%% file: main.tex
\documentclass[journal]{IEEEtran}

\usepackage{amsmath} 
\usepackage[utf8]{inputenc}
\usepackage[T1]{fontenc}
\usepackage{amsmath,amssymb,amsthm}
\usepackage{graphicx}
\usepackage{subcaption}
\usepackage{algorithmic}
\usepackage{algorithm}
\usepackage{booktabs}
\usepackage{array}
\usepackage{cite}
\usepackage{hyperref}
\usepackage{xcolor}
\usepackage{tabularx}
\usepackage{multirow}
\usepackage{balance}
\usepackage{tikz}
\usepackage{comment}
\usepackage{dsfont}
\usetikzlibrary{arrows.meta, positioning, shapes.geometric, patterns, decorations.pathreplacing}

\hypersetup{
    colorlinks=true,
    linkcolor=black,
    citecolor=black,
    urlcolor=blue
}

\newcommand{\clamp}[3]{\left[#1\right]_{#2}^{#3}}
\newcommand{\Prob}{\mathbb{P}}

\newcommand{\added}[1]{{\color{red!75!black}#1}}
\begin{document}

\title{A Theory of Probabilistic Power Provisioning for Data Centers with Distributed Energy Storage}

\author{
\IEEEauthorblockN{C. Emre Koksal\IEEEauthorrefmark{1}\IEEEauthorrefmark{3},
Richard A. Barry\IEEEauthorrefmark{2}, and
Artun Sel\IEEEauthorrefmark{1}}

\IEEEauthorblockA{\IEEEauthorrefmark{1}The Ohio State University, Columbus, OH, USA}

\IEEEauthorblockA{\IEEEauthorrefmark{2}Visiting Scientist, Research Laboratory of Electronics, Massachussetts Institute of Technology, Cambridge, MA, USA}

\IEEEauthorblockA{\IEEEauthorrefmark{3}Corresponding author. Email: koksal.2@osu.edu}
}

\maketitle

\begin{abstract}
The growing power demands and variability of AI workloads make electrical power delivery a critical constraint in data-center operation. Distributed energy storage offers a means to reduce the power capacity required to support stochastic loads, but its provisioning benefits depend fundamentally on the statistics and time scales of demand. This paper develops a probabilistic framework that jointly characterizes provisioned power, energy-storage capacity, and the probability of overdraw. We show that storage-assisted provisioning naturally separates into two operating regimes. In the Small Battery Region (SBR), overdraw is dominated by short-lived demand excursions and storage provides nearly linear reductions in the required power margin. In the Large Battery Region (LBR), overdraw results from sustained demand fluctuations across a longer span of time, and the required margin exhibits diminishing returns with storage. For this regime, we introduce \emph{effective power}, an analogue of effective bandwidth that captures the temporal statistics of the demand and provides an asymptotically tight characterization of the required provisioned power. We further quantify how temporal correlation and spatial aggregation affect storage requirements and statistical multiplexing gains, and extend the analysis to temporally correlated loads that exhibit multiple demand time scales. Finally, we evaluate the framework using power-demand traces from three production data centers that span HPC, GPU-training, and cloud-service workloads. Despite their heterogeneous and cyclo-stationary multi-modal behavior, the measured workloads exhibit the predicted operating regimes. In addition, we show a simple 4 parameter two-state model captures the fundamental dynamics governing their storage-power tradeoffs. The resulting framework provides both a probabilistic foundation and practical dimensioning principles for storage-assisted power provisioning in next-generation AI data centers.
\end{abstract}

\input{introduction}
\input{related_work}
\input{model}

\input{NB_baseline}
\input{SBR}

\input{large_battery}
\input{stat_mux}

\input{correlated}
\input{data_analysis}
\input{conclusion}

\bibliographystyle{IEEEtran}
\bibliography{refs}
\appendices

\input{MC}
\input{prop_eff_pow}
\input{typical_overdraw}
\input{app-region2a}

\input{Appendix_setting_model_parameters}

\balance
\end{document}

%% file: introduction.tex
\section{Introduction}
\label{sec:introduction}

Data centers provision a fixed power capacity to accommodate worst-case demand, resulting in significant under-utilization~\cite{barroso2019}. The natural evolution of workloads and hardware upgrades pushes the power margin to a substantial fraction of the mean demand, and the system becomes \emph{oversubscribed}: demand can, on rare occasions, exceed the provisioned power (an event referred to as \emph{overdraw}). Oversubscription is standard practice~\cite{sakalkar2020, fan2007}, and operators employ various techniques mainly on the demand side~\cite{thunderbolt2020, kumbhare2021, hsu2018smoothoperator, coach2025}, including power shaving and capping, job scheduling, workload migration, etc. to manage the risk of overdraw while keeping its probability below a prescribed target. The supply side, by contrast, has largely been treated as static: a fixed power capacity determined at design time. While energy storage devices provide intermittent and variable power availability to serve the jobs, in traditional designs, they have been mainly employed as a backup to eliminate service cuts during power outages~\cite{kontorinis2012}.
Recent measurements also document the abrupt transitions of AI workloads between high and low power regimes~\cite{li2024unseen}, while grid interconnection capacity has become a binding constraint on data center growth~\cite{saeed2026georgia}.



In this paper, we utilize the added degree of freedom of energy storage to meet statistical overdraw constraints with reduced power margins. Despite the practical relevance of energy-storage-assisted power management, a statistical framework to evaluate the tradeoff between storage capacity and required power margin has been lacking, with prior work relying mainly on simulation~\cite{fan2007, kontorinis2012}, convex or dynamic optimization~\cite{urgaonkar2011, dabbagh2019, goiri2015}, or heuristic provisioning and control~\cite{wang2012,
aksanli2013, sun2018hier}. To address the gap, we develop a rigorous probabilistic framework to study how energy storage reduces the required power margin while maintaining a statistical guarantee on minimal overdraws. To analyze the required margin, the demand is modeled by a probabilistic model that has variations in multiple time scales, which we then analyze using various probability-theoretic techniques commonly used in data network queuing theory. After building the foundations, we apply and validate our theory using three real-world data sets.

Our focus here is on a single \textit{energy group}, consisting of multiple servers sharing a dedicated power supply and battery, but our framework and results also provide architectural guidance for integrating storage devices across multiple groups at different \textit{layers} (servers, racks, rows, power distribution units (PDUs), etc.). The closest architectural precursor in the literature is~\cite{sun2018hier}, which proposes a hierarchical hybrid storage scheme pairing batteries at the row level with supercapacitors at the rack level, but uses heuristic budget allocation without statistical guarantees on overdraw. Based on our analysis here, the dynamics of the load naturally decouple into distinct time-scales, with longer term variation, e.g. cyclo-stationary daily cycles, being handled by larger batteries at higher layers, and with faster time dynamics being smoothed out with smaller batteries at the lower layers. 

Our core insights suggest that storage-assisted power systems operate in two distinct regions, based on the battery size measured relative to the demand statistics:
\begin{itemize}
    \item In the \emph{small battery region (SBR)}, the battery is almost always full and overdraws are driven by single-slot excess demand spikes followed by a rapid recharging of the battery. In the SBR, power savings are nearly optimal and linear with storage size with each unit of storage saving approximately one unit of provisioned power. In this region, the battery quickly transitions from full to empty and then rapidly recharges.
    \item In the \emph{large battery region (LBR)}, most likely overdraws occur when the battery was already partially drained by previous sustained multi-slot excess demands. In the LBR, the power savings exhibit diminishing returns with the relationship between storage size and margin more complex than in the SBR, exhibiting in general multiple battery sub-regions.
\end{itemize}

Each region's size, its potential margin reduction, and the transition between regions are all primarily governed by the \textit{demand statistics}, and much less so on the overdraw guarantees. For example, if the demands are well modeled as a single state Gaussian without significant temporal correlations, then the SBR accounts for roughly half of the total achievable margin reduction with the LBR achieving the other half. For multi-state demand statistics, the SBR can achieve nearly the same result as measured with respect to the highest demand periods, but far less a percentage of the overall potential margin reduction.

Core of dimensioning a multi-layer system is to quantify the effects of statistical multiplexing on margin and storage requirements of a group. It is well known and previously studied that aggregating a larger number of servers into a group can dramatically reduce the required margin/server due \textit{to spatial} statistical multiplexing\added{~\cite{fan2007}}.  Here, we quantify the ability of energy storage to reduce the margin/server due to \textit{temporal} statistical multiplexing, with 
batteries providing larger gains due to the ability to average demands over extended periods of time.  However, larger groups also require larger batteries, and thus the trade offs of larger vs. more numerous smaller groups is more complex. We find that the combination of the two forms can be additive, or in some cases neutral or even mildly subtractive, depending on how energy storage scales with group size.

Key to our analysis, we introduce the concept of \textbf{effective power}, based on the concept of effective bandwidth used in network theory \cite{kelly1991}. We show that effective power is an accurate predictor of the required margin for both single state i.i.d. processes as well as more complex multi-state demand processes for larger battery sizes. As part of this analysis, we quantify the typical time it takes to drain the battery to overdraw, with larger batteries having longer typical overdraw times. For small batteries, and thus short overdraw times, the effective power is less useful and we rely on simple yet tight bounds, Markov analysis, and simulations.

To summarize, \textbf{contributions of this paper} span both conceptual  and practical aspects of data-center design. On the theoretical side, we develop a probabilistic framework that characterizes the tradeoff between provisioned power, distributed energy storage, and statistical overflow guarantees. The framework introduces \textbf{effective power} as the natural analogue of effective bandwidth for energy provisioning, establishing linear provisioning conditions despite the underlying stochastic complexity. We identify two distinct operating regimes: the small-battery and large-battery and characterize the required power margin, the role of temporal correlations, and the scaling laws governing both spatial and temporal statistical multiplexing. We also provide new \textbf{architectural insights} for next-generation AI data centers. In particular, we demonstrate how energy storage can transform the supply side from a static design constraint into a adjustable resource, identify the battery sizes at which diminishing returns emerge, quantify the interplay between storage deployment and server aggregation, and show how different time scales of demand naturally map to hierarchical energy-storage architectures. Finally, the framework is validated using three production-scale data-center traces, demonstrating both the predictive accuracy of the proposed models and their usefulness for practical energy-system dimensioning.

The remainder of this paper is organized as follows. Section~\ref{sec:related-work} covers related work in this area. Section~\ref{sec:system-model} then introduces the system model and formal problem statement. Section~\ref{sec:baseline} calculates the required margin for a traditional system without a battery under our statistical assumptions.  Margin requirements for the small battery region are analyzed in Section~\ref{sec:SBR}, focusing mainly on i.i.d. demand processes. Section~\ref{sec:eff_bw} then introduces the concept of effective power and analyzes the margin requirements in the large battery region for a large class of demand processes. Section~\ref{sec:stat-mux} then quantifies the trade offs between spatial and temporal multiplexing for both operating regions. Section~\ref{sec:correlated} applies the previous results to a multi-state Markov modulated process. Section~\ref{sec:data sets} then compares our analysis to three real-world data sets. A summary of conclusions and future research directions are presented in Section~\ref{sec:conclusion}.

%% file: related_work.tex
\section{Related Work}
\label{sec:related-work}


Electricity demand from data centers has grown quickly with the
adoption of AI and is expected to reach a large fraction of national
electricity supply~\cite{epri2024powering}. In response, industry has
begun to redesign how data-center power is
delivered~\cite{schneiderWP110}. The problem of meeting a variable power
demand from a fixed provisioned power supply has been studied in three main
directions. The first is demand-side management, covering cluster-level
oversubscription and capping, workload scheduling and placement, and
server-level power capping. These methods control when and how much load
runs as a power limit is approached. The second is hardware energy
buffering, which places fast storage next to the compute. The third is
the analytical toolkit of effective bandwidths and large deviations,
used here to size energy storage.

\subsection{Demand-side power oversubscription and capping}
Power oversubscription provisions below the worst-case aggregate
demand while managing the residual overdraw risk. It is the main
industrial way to improve utilization. The control is almost
always on the demand side, where non-critical work is throttled at a
binding choke point so that the draw stays below the breaker rating.
A power-capping system deployed in Google's data centers which keeps draws
below the limit by throttling the CPU bandwidth of batch jobs while
protecting latency-sensitive ones, results in a reported $9$--$25\%$ gain in
oversubscription~\cite{thunderbolt2020}. In the public
cloud, where the platform cannot see inside tenant VMs, Azure's
prediction-based capping doubled the safe oversubscription level, from
about $6\%$ to about $12\%$, by learning which VMs can absorb
throttling~\cite{kumbhare2021}. This built on earlier analysis of how
much oversubscription is safe given breaker-violation
statistics~\cite{fu2011}. These methods reduce the effect of overdraw
when it occurs rather than reducing the provisioned margin with a
supply-side resource. The same work observes that storage-based peak shaving can be used in a
complementary fashion with demand-side capping rather than as a
replacement for it~\cite{kumbhare2021}, and that combination is made
precise in the present paper.

\subsection{Scheduling, placement, and power reallocation}
A second group of demand-side methods reshapes where and when
loads runs. Workload placement spreads services that peak at the same
time of day, so that no single part of the power distribution is
overloaded while other parts sit idle. This makes room for up to $13\%$
more machines on the same
infrastructure~\cite{hsu2018smoothoperator}. Power is often kept idle as a backup in case of hardware failures.
Putting that backup power to use during normal operation, and
reclaiming it when a failure occurs, makes room for up to $33\%$ more
servers~\cite{flex2021}. Because cloud workloads rarely peak at the
same time, packing together VMs whose busy periods do not overlap lets
them share resources and fit up to $26\%$ more on the same
hardware~\cite{coach2025}. Capacity can also be added by adjusting the
hardware itself. A multi-year DVFS-boosting deployment at Meta produced
about $12$~MW of effective supply~\cite{piga2024dvfs}, and idle-resource
leasing reclaimed about $25\%$ of the physical footprint at fleet
scale~\cite{gupta2024idle}. For GPU and large language model
workloads, the spare power in inference clusters has been used to host
up to $30\%$ more servers~\cite{patel2024llm}, and scheduling that
accounts for temperature as well as power reaches up to $40\%$ more
oversubscription~\cite{tapas2025}. All of these act on the demand side
and are separate from the storage layer studied in this paper.

\subsection{Server-level power capping}
At the level of a single server, power capping enforces a fixed
power budget by trading performance for safety, which on modern machines
requires coordinating several power domains. One approach uses
reinforcement learning to divide a server's power budget among its CPUs
and GPUs while keeping throughput high~\cite{powercoord2020}. Jointly
tuning the CPU and GPU clock speeds, rather than capping just one, gives
better machine-learning inference under the same power
budget~\cite{ma2025capgpu}. The same controls can also
be used to save energy during training. Adjusting the GPU power limit
together with the batch size lowers the energy needed to train a neural
network~\cite{you2023zeus}, and setting the GPU clock speed for each
stage of a large training job removes wasted
energy~\cite{chung2024perseus}. These in-server controls are
complementary to the between-server storage layer studied in this
paper. They set how a power cap is enforced, while distributed storage
sets how much that cap can be relaxed.

\subsection{Hardware energy buffering and storage technology}
Most closely related to the supply-side approach of this paper is
the use of energy storage to manage power capacity in data centers,
which has recently been surveyed for AI workloads across battery,
supercapacitor, and hybrid storage technologies and
timescales~\cite{rahman2026ess}. The idea predates the AI era. Hybrid
battery and supercapacitor stores have been used to shave and cap
data-center peak power, trading energy store for provisioned power to
lower both capital and operating cost~\cite{zheng2017hybridess}. Recently the approach has reached commercial products such as NVIDIA's
GB300 NVL72 rack, which places energy storage next to the GPUs (about
$65$~joules per GPU) that charges when the GPUs are idle and discharges
during bursts, lowering the average peak grid demand by about
$30\%$~\cite{nvidia_gb300}. Supercapacitors are also
being used as millisecond-scale buffers that separate the bursty power
demand of AI clusters from the stability needs of the
grid~\cite{genkina2025supercap}. 
Battery storage has also been used to enforce grid ramp rate limits for AI data centers through online control~\cite{abera2026battery}, and workload deferral or spatial shifting offers a complementary demand side lever that reduces the required interconnection power~\cite{chen2026defer}.
On the device side, lithium
iron-phosphate chemistries now dominate stationary
storage~\cite{chen2024lfp}, and supercapacitor technology continues to
improve in energy density~\cite{salaheldeen2025supercap}. These works
show that storage is a practical way to manage power capacity, but they
only consists of specific engineering solutions and heuristics. None of it
gives a quantitative theory for how large a store must be to meet a
stated overdraw probability at a given margin reduction.

\subsection{Effective bandwidths and large deviations}
The tools used here come from a body of work on effective
bandwidths and large deviations developed for communication networks.
An effective bandwidth summarizes a bursty source by a single rate
between its mean and its peak, which makes it possible to decide how
many sources a link can carry~\cite{kelly1996notes, courcoubetis1995,
deveciana1995effective}. Related results estimate the probability that a
queue exceeds a large level, both for the waiting time~\cite{abate1995}
and for systems with many busy servers~\cite{whitt1984}, and they
describe how sources that vary over several time scales combine when
multiplexed~\cite{tse1995multiplexing, grossglauser2003}. The same kind
of statistical characterization has also been applied to other
time-varying resources, such as routing over wireless links whose
quality changes in time~\cite{koksal2006qaware}. These tools were
developed for communication systems rather than for sizing energy
storage.

Across all of the above directions, no prior work gives a closed-form
probabilistic description of the tradeoff between the
provisioned-power margin and the size of distributed energy storage
under realistic demand.

%% file: model.tex
\section{System Model}
\label{sec:system-model}

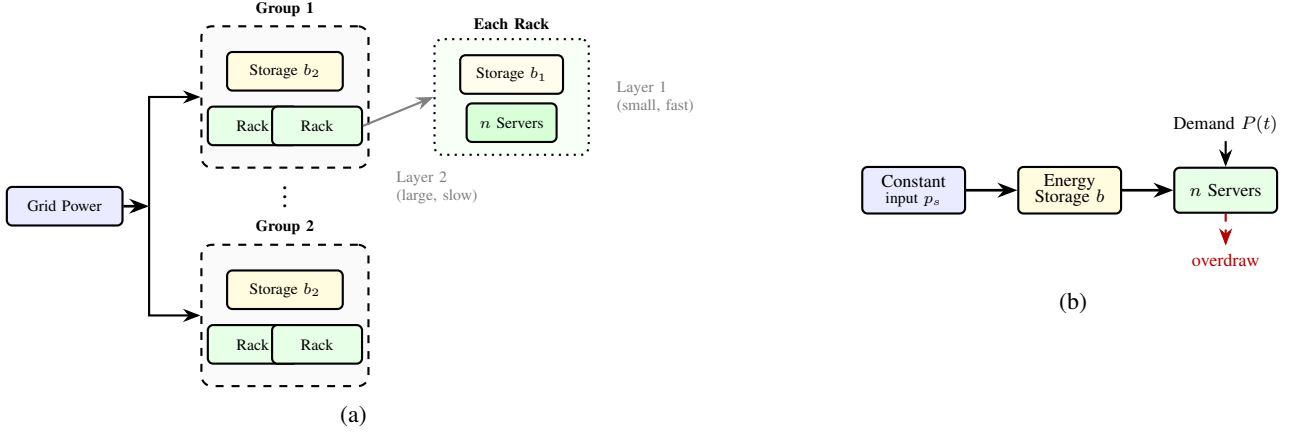
\begin{figure*}[t]
    \centering
    \begin{subfigure}[c]{0.55\textwidth}
    \centering
    \begin{tikzpicture}[
        >=Stealth, scale=0.85, transform shape,
        block/.style={draw, thick, minimum width=1.4cm, minimum height=0.6cm, align=center, rounded corners=2pt, font=\scriptsize},
        arr/.style={->, thick},
        every node/.style={font=\scriptsize}
    ]
    \node[block, fill=blue!8, minimum width=1.8cm] (grid) {Grid Power};

    \node[draw, thick, dashed, rounded corners=4pt,
          minimum width=2.6cm, minimum height=2.2cm,
          right=1.2cm of grid, yshift=1.7cm, fill=gray!4] (g1) {};
    \node[block, fill=yellow!15, minimum width=1.8cm] at ([yshift=0.4cm]g1.center) (b1) {Storage $b_2$};
    \node[block, fill=green!10] at ([yshift=-0.45cm, xshift=-0.5cm]g1.center) (r1a) {Rack};
    \node[block, fill=green!10] at ([yshift=-0.45cm, xshift=0.5cm]g1.center) (r1b) {Rack};
    \node[above=0.02cm of g1.north, font=\scriptsize\bfseries] {Group 1};

    \node[draw, thick, dashed, rounded corners=4pt,
          minimum width=2.6cm, minimum height=2.2cm,
          right=1.2cm of grid, yshift=-1.7cm, fill=gray!4] (g2) {};
    \node[block, fill=yellow!15, minimum width=1.8cm] at ([yshift=0.4cm]g2.center) (b2) {Storage $b_2$};
    \node[block, fill=green!10] at ([yshift=-0.45cm, xshift=-0.5cm]g2.center) (r2a) {Rack};
    \node[block, fill=green!10] at ([yshift=-0.45cm, xshift=0.5cm]g2.center) (r2b) {Rack};
    \node[above=0.02cm of g2.north, font=\scriptsize\bfseries] {Group 2};

    \node at (g2.north) [above=0.45cm, font=\large] {$\vdots$};

    \draw[arr, line width=1pt] (grid.east) -- ++(0.4,0) coordinate (split);
    \draw[arr, line width=0.8pt] (split) |- (g1.west);
    \draw[arr, line width=0.8pt] (split) |- (g2.west);

    \node[draw, thick, dotted, rounded corners=3pt,
          minimum width=2.4cm, minimum height=1.8cm,
          right=1.0cm of g1, yshift=0cm, fill=green!3] (rd) {};
    \node[block, fill=yellow!8, minimum width=1.6cm] at ([yshift=0.35cm]rd.center) (b2s) {Storage $b_1$};
    \node[block, fill=green!15] at ([yshift=-0.4cm]rd.center) (srvs) {$n$ Servers};
    \node[above=0.02cm of rd.north, font=\scriptsize\bfseries] {Each Rack};

    \draw[arr, line width=0.8pt, gray] (r1b.east) -- (rd.west);

    \node[right=0.3cm of rd, font=\scriptsize, align=left, text=gray] {Layer 1\\(small, fast)};
    \node[right=0.3cm of g1.east, yshift=-1.4cm, font=\scriptsize, align=left, text=gray] {Layer 2\\(large, slow)};

    \end{tikzpicture}
    \caption{}
    \label{fig:layered-arch}
    \end{subfigure}
    \hfill
    \begin{subfigure}[c]{0.40\textwidth}
    \centering
    \begin{tikzpicture}[
        >=Stealth, scale=0.85, transform shape,
        block/.style={draw, thick, minimum width=1.6cm, minimum height=0.7cm, align=center, rounded corners=2pt, font=\footnotesize},
        arr/.style={->, thick},
        every node/.style={font=\footnotesize}
    ]
    \node[block, fill=blue!8] (input) {Constant\\[-2pt]\scriptsize input $p_s$};

    \node[block, fill=yellow!15, right=0.8cm of input] (storage) {Energy\\[-2pt]Storage $b$};

    \node[block, fill=green!10, right=0.8cm of storage] (servers) {$n$ Servers};

    \node[above=0.4cm of servers] (demand) {Demand $P(t)$};
    \draw[arr] (demand) -- (servers);

    \draw[arr, line width=1pt] (input) -- (storage);
    \draw[arr, line width=1pt] (storage) -- (servers);

    \draw[arr, red!70!black, dashed, line width=0.8pt] (servers.south) -- ++(0,-0.5)
        node[below, red!70!black] {overdraw};

    \end{tikzpicture}
    \caption{}
    \label{fig:single-layer}
    \end{subfigure}

    \caption{(a) 2 Layer architecture: Each Layer~2 groups has storage $b_2$; Layer 1 groups have storage $b_1$. (b) The single-layer abstraction used throughout this paper applies at each layer independently.}
    \label{fig:system-diagram}
\end{figure*}

\noindent This section formalizes the system model, defines the key quantities, and states the central problem. 

\subsection{System Description}

We consider a hierarchical architecture such as shown in Fig.~\ref{fig:layered-arch}, where each layer consists of multiple \textit{energy groups}, each with its own \textit{energy storage system}.  For simplicity, we refer to each group's energy storage system as a \textit{battery}, but a system could consist of multiple devices. Different groups, and in particular different layers, will have different performance requirements and thus may utilize different technologies. For example, at the rack level, fast super capacitors operating on the millisecond timescales may be appropriate, whereas higher layer groups may utilize slower but higher energy density technologies such as Lithium-ion or lead-acid batteries. Each energy system could be built from a single technology or could combine heterogeneous technologies optimized for different tradeoffs across storage, speed, cost, etc.~\cite{sun2018hier, salaheldeen2025supercap}. 

Our focus in this paper is a single energy group at the lowest layer as shown in Fig.~\ref{fig:single-layer}. A given group could represent a single server or a group of servers in one or more racks. The $n$ servers of a group share a constant power supply of size $p_s$ and battery of maximum energy capacity $b_e$.  When the group's power demand exceeds the power supply $p_s$, the servers can draw on the battery to avoid a power \textit{overdraw}. Similarly, when the demand is less than the supply, the battery can be charged up to its maximum storage capacity $b$. We assume that power demands cannot be delayed, and that each group operates independently and uses its storage greedily, charging/discharging its battery based on instantaneous demand. 

While we focus on the lowest-layer energy group, the results here are applicable to higher layers, where $n$ would represent the number of energy groups in the layer below it. Note, however, that factors involving interactions across layers are not addressed in this paper. Examples include lower layer storage smoothing out power demand variations at higher layers, that those demands may interact with each other through shared power supplies, that batteries may be utilized in ways that are not purely greedy, i.e., a group may balance its own needs versus those of its neighbors, and the effects of power management and job orchestration algorithms.

\subsection{Energy Group Model}

The energy group shown in Fig.~\ref{fig:single-layer} is modeled as a time slotted system of length $T_s$ seconds, where time slot $t$ represents physical time $(t-0.5)T_s$ to $(t+0.5)T_s$. The model has 3 components: the power supply, the individual and group power demands, and the energy storage.

\begin{itemize}
    \item \textbf{Power Supply}: The supply is modeled as a constant $p_{s}$ units of power available in every slot, in Watts.
    \item \textbf{Power Demands}: Each group consists of $n$ \textit{individual power demands}, modeled as a stationary discrete time stochastic process, $X_i(t) \geq 0$, where $X_i(t)$ represents the average power demand during slot $t$ in Watts. The \textit{group power demand} $P(t)$ is the sum of the individual demands, representing the average power demand of the group during slot $t$, 
\begin{align}
     P(t) =  \sum_{i=1}^{n} X_i(t)
    \label{eq:group-power-demand}
\end{align}
\item \textbf{Energy Storage}: The battery is modeled as an energy storage device of size $b_e \geq 0$ joules. In time slot $t$, the battery starts with \textit{energy} $B_e(t)  \in [0,b_e]$ indicating that it can provide up to $B_e(t)/T_s$ Watts of power in slot $t$. $B_e(t) = 0$ indicates an empty battery, and $B_e(t) = b_e$ indicates a full battery. The battery has the potential to fully charge or discharge in a single slot, with the actual amount of discharge/charge determined by the excess/deficiency energy demand over the supply during that slot. Specifically, the evolution of $B_e(t)$ is given by the \textit{battery energy state equation}
\begin{equation}   
B_e(t+1) = \left[ p_{s}T_s + B_e(t) - P(t)T_s \right]_0^{b_{e}}
    \label{eq:battery-update-energy}
\end{equation}
where $\clamp{z}{0}{b_e} = \min(b_e, \max(z, 0))$ denotes the clamping operator, and $B_e(t+1)$ is both the total energy available at the end of slot $t$ and the start of slot $t+1$.
\end{itemize}

Since time slots are of constant fixed length, it is possible and more convenient to rewrite the energy state equation in terms of power by dividing each side by $T_s$ and defining $B(t) = B_e(t)/T_s$ to be the \textit{power charge} of the battery, representing the amount of power, in Watts, the battery can supply during time slot $t$. For instance, if a time slot size was 30 seconds, each watt of power charge is equivalent to 30J = $1/120$ kWh of battery energy.  Similarly defining $b = b_e/T_s$ to be the \textit{power charge capacity} of the battery, we have the \textit{power state equation}
\begin{equation}
    B(t+1) = \left[ p_{s} + B(t) - P(t)\right]_0^b
\label{eq:battery-update}
\end{equation}
By measuring the battery storage in units of power charge, and using the power state equation, we can investigate the performance of the system (Fig. \ref{fig:single-layer}) independent of the time slot length. In plots, by convention, we convert the battery power charge to energy using a specified slot length, or more generally in units of watt-slots using $T_s = 1$ slot.  To convert to joules for any particular physical system, multiply either by the actual slot length in seconds. 

Note that for ease of explanation we have chosen watts and joules as the base units. In addition, given the current demands of servers, this unit makes $B(t)$ large enough to be well approximated as discrete and only taking on positive integer values. We can thus move freely between continuous and discrete demand processes, e.g. Gaussian and Poisson, as convenient. Appendix \ref{app:markov-discretization} describes more details on these issues for numerical calculations and simulations.

\subsection{Statistical Assumptions}

Battery state evolution, Eq.~(\ref{eq:battery-update}), applies to any sample process $P(t)$, including the real data sets used to evaluate our model. In this case, since the data sets are finite, we take $t=0$ as the starting point and $B(0)$ as the starting battery charge. The issues associated with choosing $B(0)$ are discussed in Section~\ref{sec:data-drive-simulations}.

For statistical models, we assume the system has been running forever and that the individual demand processes, $X_i(t)$, are all statistically identical and are 1) stationary, 2) ergodic, and 3) non-negative and finite. It follows then that they all have log moment generating functions (LMGF) over some finite region around $0$ of the transformation variable $r$. Although not all distributions have LMGFs, the fact that real world power demand is positive and finite implies that any reasonable probabilistic model for it will have one over a cyclo-stationary period of interest.  Note that the assumptions do not preclude demand processes with temporal correlations, although that will be a simplifying assumption in parts of the paper. Beyond the above, the statistics of $X_i(t)$ are arbitrary. Note that when modeling a particular physical system, the statistics of $X_i(t)$ are dependent on the slot length $T_s$, with longer slot lengths tending to smooth out $X(t)$.

Given above, each process $i$ is an instance of a stationary, ergodic, discrete time process $X(t)$, where for any $t$, the random variable $X(t)$ is characterized by
\begin{align}
\mu_X & = \mathbb{E}[X(t)] \\
\sigma^2_X & = \mathbb{E}[(X(t) - \mu_X)^2] \\
\Lambda_X(r) & = \log \mathbb{E}[e^{r \space X(t)}] \\
\bar{F}_X(d) & = \mathbb{P}[X(t) > d] \\
\bar{F}^{-1}_X(\varepsilon) &= \min_d\{\bar{F}_X(d) \leq \varepsilon\}
\end{align}
which are its mean, variance, LMGF, complementary cumulative distribution function, and quantile functions respectively, and $\mathbb{E}[\cdot]$ and $\mathbb{P}[\cdot]$ denote expectation and probability. $\bar{F}_X(d)$ is called $X$'s survivor function or tail probability.

If follows from above, that the group power is a stationary, ergodic, non-negative finite random process $P(t)$, where for any $t$, the random variable $P(t)$ is characterized by
\begin{align*}
\mu_P & = \mathbb{E}[P(t)] = n \mu_X \\
\sigma_P^2 & = \mathbb{E}[(P(t) - \mu_P)^2] \\
\Lambda_P(r) & = \log \mathbb{E}[e^{r P(t)}] \\
\bar{F}_P(d) & = \mathbb{P}[P(t) > d] \\
\bar{F}^{-1}_P(\varepsilon) &= \min_d\{\bar{F}_P(d) \leq \varepsilon\}, 
\end{align*}
which are its mean, variance, LMGF, tail, and quantile functions, respectively. Let $p_{\max}$ denote the maximum value of $P(t)$, if it exists\footnote{The power drawn in real loads is always upper bounded by a finite amount. However, typical models used for the load (such as Poisson or Gaussian) can be unbounded. For such models, $p_{\max}$ can be set to to a point in the tail beyond the overdraw target rate}. In the special case where the individual demands are statistically independent, $\sigma_P^2 = n \sigma_X^2$ and $\Lambda_P(r) = n \Lambda_X(r)$. 

Finally, define the asymptotic LMGF of $P(t),\ \bar{\Lambda}_P(r)$,
\begin{align}
    \bar{\Lambda}_P (r)&\triangleq \lim_{T \rightarrow \infty} \frac{1}{T} \log \mathbb{E}\left[ \exp \left(r\sum_{t=1}^T   P(t) \right)\right]. \label{def:asymLMGF}
\end{align}
We will use the asymptotic LMGF to evaluate and bound the tail probabilities of the cumulative demand process over multiple slots.

\subsection{Overdraw, Capacity, and Margin}

An \textit{overdraw} occurs at time $t$ if the group power demand $P(t)$ exceeds the available power, whether supplied or stored. Since $P(t)$ is stationary and ergodic, we can define the probability of an overdraw in any arbitrary time slot, 
\begin{align}
   \mathbb{P} [overdraw] & = \mathbb{P}\left(P(t) > p_{s}  + B(t)\right)
    \label{eq:overdraw-rate}
\end{align}
and note that this probability is also the long term \textit{overdraw rate} per slot, defined as the fraction of slots which overdraw.

When overdraws are rare, the average power consumed by the group is approximately the average demand, $\mu_P$, with or without a battery.  However, batteries smooth out variations in consumed power, and thus smooth out the power demand of a higher layer in a multi-layer system like Fig.~\ref{fig:layered-arch}. Since data center operators regularly utilize demand side power management techniques such as power shaving, an overdraw event in our model represents the time that demand must be curtailed. The insights provided in this paper can be used to inform such strategies and/or jointly optimize the supply and demand sides.

To that end, the central question addressed in this paper is how much power, supplied or stored, is required to keep the overdraw rate below a low target value $\varepsilon$, and secondarily how it changes with the number of servers in a group.  To that end, we define the \textit{required power} $p_r(b)$ to be the minimum supplied power to achieve that goal with a battery size of $b$, and the \textit{required margin} to be the required power over the average demand,
\begin{align}
    p_r(b) &= \min\!\big\{p_{s} : \mathbb{P}[overdraw] \leq \varepsilon\big\}
    \label{eq:required-capacity} \\
    m(b) &= p_r(b) - \mu_{p} \label{eq:required-margin}
\end{align}
where $p_r(0)$ and $m(0)$ represents the no battery required supply power and margin, respectively. Note that the required power and margin are both functions of the target overdraw $\varepsilon$ as well as the statistical characteristics of $P(t)$. Since a system with a battery will perform at least as well as a system without one, and larger batteries cannot do worse than smaller batteries, we have that $p_r(b)$ and $m(b)$ are both non-increasing functions of $b$. Further, as $b \rightarrow \infty$, the required power decreases to the average demand $\mu_P$, and the required margin to $0$.

\subsection{Choke-Point Utilization and Oversubscription Ratio}

The \textit{operating oversubscription ratio OSR} is the amount of excess provisioned IT equipment power, $p_{\max}$, over the supplied power, $p_s$, i.e. $\text{OSR} = p_{\max} / p_s$. \cite{sakalkar2020,fu2011}. Here $p_{\max}$ could be the total nameplate power rating of the group's servers, a measured value of such, or a contractual agreement on such in a multi-tenant environment. Very often OSR is expressed at a \%, i.e. $\text{OSR} \% = (\text{OSR} - 1)\times 100$.

The higher the OSR (more IT equipment or less deployed power), the more efficient the data center. However, the system cannot operate at an arbitrary OSR without risking power hardware overloads and circuit breaks. To maintain acceptable performance, the system must operate below some maximum OSR, usually measured at some choke point in the data center. In our terminology,
\begin{align}
\mathrm{OSR_{\max}}(b) & = \frac{p_{\max}}{p_r(b)} = \frac{p_{\max}}{\mu_P + m(b)}.
\end{align}
Keeping the operating OSR below the maximum OSR guarantees the system will not overload more than the target overdraw rate. Since energy storage reduces the required margin (at some level of system performance), a combined power/energy group allows the data center to operate a higher OSR, reducing costs.

Since the consequences of a hardware overload are severe, potentially bringing down one or more groups, power capping techniques can be introduced which maintain the drawn power below some limit, generally set some safe distance below an overload. Since overdraws have less consequences than overloads, the target overdraw rate can be set higher, thus allowing higher $\text{OSR}_{\max}$. For instance, it may be acceptable to operate at daily short term overdraws, but not overloads.

Another relevant related metric is $p_s / p_r(b)$, which represents the amount of excess power deployed over what is required and is a measure of the untapped potential of the system which could be exploited by deploying more equipment or reducing power.   The amount of this potential due to energy storage is $p_r(0)/p_r(b)$.

The complete notation used throughout this paper is summarized in Table~\ref{tab:notation}.

\begin{table}[t]
\centering
\caption{Summary of Notation}
\label{tab:notation}
\renewcommand{\arraystretch}{1.15}
\footnotesize
\begin{tabular}{@{}clc@{}}
\toprule
\textbf{Symbol} & \textbf{Description} & \textbf{Unit} \\
\midrule
$X(t)$ & Per-server power demand in slot $t$ & W\\
$\mu_X$ & Mean per-server demand & W\\
$\sigma_X^2$ & Variance of per-server demand & W$^2$\\
$\Lambda_X(r)$ & Log moment generating function of $X(t)$ & --- \\
 $\bar{F}_X(d)$& CCDF of X, $\mathbb{P}[X(t) > d]$&--- \\
 $\bar{F}^{-1}_X(\varepsilon)$& $\varepsilon$-quantile of the server demand&W\\
$n$ & Number of servers in a group & --- \\
\midrule
$P(t)$ & Group power demand in slot $t$ & W\\
$\mu_P$ & Mean group demand & W\\
$\sigma_P^2$ & Variance of group demand & W$^2$\\
 $p_{\max}$ & Maximum group equipment power&W\\
$\Lambda_P(r)$ & Log moment generating function of $P(t)$ & --- \\
$\bar{\Lambda}_P(r)$ & Asymptotic (per-slot) LMGF of $P(t)$ & --- \\
$\sigma_\text{eff}^2$ & Effective (long-run) variance of $P(t)$ & W$^2$\\
$\bar{F}_P(d)$ & Group demand tail: $\mathbb{P}[P(t) > d]$ & --- \\
$\bar{F}^{-1}_P(\varepsilon)$ & $\varepsilon$-quantile of the group demand & W\\
\midrule
$p_s$ & Supplied power & W\\
$\mathrm{OSR}$ & Oversubscription ratio: $p_{\max}/p_s$ & --- \\
$T_s$ & Slot duration & s \\
 $b_e$ & Battery energy capacity: $T_s\, b$ &J\\
$b$ & Battery power charge capacity: $b_e/T_s$& W\\
$B(t)$ & Battery power charge at start of slot $t$ & W\\
\midrule
$\varepsilon$ & Maximum allowable overdraw rate & --- \\
$\delta$ & Overdraw exponent target: $-\log \varepsilon$ & --- \\
$\alpha$ & Gaussian quantile coeff.: $\sqrt{2\delta - \ln(4\pi\delta)}$ & --- \\
\midrule
$p_r(b)$ & Required power with storage $b$ & W\\
$p_r(0)$ & No-storage required power: $\bar{F}^{-1}_P(\varepsilon)$ & W\\
$m(b)$ & Required margin: $p_r(b) - \mu_P$ & W\\
$m(0)$ & No-storage margin & W\\
$\tilde{m}(b)$ & Normalized margin: $m(b)/m(0)$ & --- \\
$\tilde{b}$ & Normalized battery size: $b/m(0)$ & --- \\
$q_n(b)$ & Margin per server: $m(b)/n$ & W\\
\midrule
$b_\text{tx}$ & SBR/LBR transition battery size in power charge& W\\
$\beta_\text{tx}$ & Normalized transition point: $b_\text{tx}/m(0)$ & --- \\
$b_{\text{tx},i}$ & Region transition points ($i=1,2,3$) & W\\
\midrule
$\pi_k$ & Stationary probability of demand state $k$ & --- \\
$L_k$ & Mean dwell time of state $k$ ($L$: common value) & slots \\
$\mu_k,\ \sigma_k$ & State-$k$ demand mean, std.\ dev. & W\\
$\delta_1,\ \alpha_1$ & HIGH-state $\delta$ and $\alpha$ (also $\delta_H,\alpha_H$) & --- \\
\midrule
$r^*$ & Saddle-point tilting parameter & (W-slot)$^{-1}$\\
$P_\text{eff}(r)$ & Effective power: $\bar{\Lambda}_P(r)/r$ & W\\
$T_\text{typ}(b)$ & Typical overdraw duration & slots \\
$I_P(p_s)$ & No-battery (NBR) rate function & --- \\
$I_B(p_s)$ & LBR overdraw decay-rate function & ---\\
\bottomrule
\end{tabular}
\end{table}

%% file: NB_baseline.tex
\section{No-battery Regime (NBR) baseline}
\label{sec:baseline}

This section analyzes the required power $p_r(0)$ and margin $m(0)$ required to meet the overdraw target~$\varepsilon$ without energy storage. This serves as our baseline for the benefits of spatial and temporal multiplexing.

Without a battery, the overdraw probability depends solely on the tail of the demand distribution in a given slot, i.e. 
\begin{align}
   \mathbb{P}[overdraw] &= \mathbb{P}\left( P(t) > p_s \right) \\
   &= \bar{F}_P(p_s)
    \label{eq:NB-overdraw-rate}
\end{align}
Thus, the no-battery required power and margin are functions of the demand distribution's quantile function,
\begin{align}
\label{eqn:cap-NB}
p_r(0) &= \bar{F}^{-1}_P(\varepsilon) \\
\label{eqn:margin-NB}
m(0) &= \bar{F}^{-1}_P(\varepsilon)  - \mu_P
\end{align}
Here, $m(0)$ represents both the required power margin to meet the overdraw target, as well as the \textit{maximum tolerable slot burst} of the group power demand over the mean, i.e. bursts larger than $m(0)$ occur with less than $\varepsilon$ probability and are deemed acceptable. 

Eq. (\ref{eqn:cap-NB}-\ref{eqn:margin-NB}) apply equally well to $P(t)$ with or without temporal correlations and are easily calculated given empirical data or a statistical model. Next, we characterize the no-battery baseline and provide examples based on insightful statistical models.

\subsection{Estimate of No-Battery Baseline}


We first provide a general result based on tail bounds. For a given log moment generating function $\Lambda_P(r)$, the Chernoff bound on the tail can be written as:
\begin{equation}
\label{eq:chernoff_no_battery}
\mathbb{P} [overdraw] \leq e^{-I_P(p_{s})} \end{equation}
where $I_P(p_s)$ is the rate equation
\begin{equation}
    I_P(p_s) = \sup_{r > 0} \big\{r p_s - \Lambda_P(r)\big\},
    \label{eq:rate-function}
\end{equation}
Since the Chernoff bound is asymptotically tight in the tail, we have that for small overdraw targets,
\begin{align}
m(0) \approx I_P^{-1}(\delta) - \mu_P
\label{eqn:margin-NB-general}
\end{align}
where $\delta = \ln{1/\varepsilon}$. Taking a second order Taylor series expansion on $\Lambda_P(r) \approx \mu_Pr + \frac{1}{2}\sigma_P^2 r^2$, 
\begin{equation}
\label{eqn:NB-margin-2ndOrder}
m(0) \approx \sigma_P \sqrt{2 \delta}
\end{equation}
showing that no-battery margin is primarily determined by the standard deviation of the group power demand when the second order Taylor expansion is accurate. 

If $P(t)$ has a distribution with a simple analytical quantile expression, then we can often derive tighter approximations. For example, for Gaussian demand with not too small a mean, using Mill's ratio~\cite{billingsley1995probability}), we can write:
\begin{align}
m(0) \approx \alpha \sigma_P,
\label{eqn:NB-margin-gaussian}
\end{align}
where $\alpha = \sqrt{2\delta - \ln(4\pi\delta)}$, and . For $\varepsilon = 10^{-3}$, $\alpha \approx 3.05$ and the the system requires about $3$ standard deviations of power demand above the average power.  The looser target of $10^{-2}$ requires about $2.27 \sigma_P$ of power margin, and a stricter demand of $\varepsilon = 10^{-6}$ requires $4.8 \sigma_p$.

In this paper, we will use three prototypical examples for group demand: Gaussians with mean 1kW and standard deviations of 32W, 79W, and 158W, respectively. For an overdraw target rate of  $\varepsilon = 10^{-3}$, the no-battery margin requirements are $\approx 98W, 241W$, and $481W$, respectively. 

Note that even for the noisiest case, 0W is more than 6 standard deviations away from the mean and thus we can safely ignore the possibility that $P(t)<0$ in some time slot. For smaller means, the Poisson distribution can be used as a reasonable model of the power demand in a slot.

For distributions with skew, a 3rd order expansion on the Taylor series can be used, showing that tails heavier than a Gaussian require a higher margin, and those with lighter tails less. For multi-modal distributions, Eq.~(\ref{eqn:NB-margin-gaussian}) can provide widely inaccurate results. That case is handled next.

\subsection{Multi-modal Distributions \label{sec:multi-mode-NB}}
 
There are important situations for which the distribution of $P(t)$ has multiple modes, e.g. corresponding to periods of high, medium, or low loads.  In these cases, the Chernoff bound as applied above can be loose in non-asymptotic situations. For these cases, we derive an alternate approximation to $m(0)$.

If $P(t)$ is multimodal, it can often be well approximated as a sum of distributions, 
\begin{equation}
f_P(d) \approx \sum_{i=1}^K a^{(i)} f_{P^{(i)}}(d)
\end{equation}
where each $f_{P^{(i)}}$ is the pdf/PMF describing a mode of the distribution, with mean $\mu_{P^{(i)}}$ and standard deviation $\sigma_{P^{(i)}}$, and $a^{(i)}$ is the weight of each mode. Without loss of generality, order the states so the means are decreasing. We refer to state 1 as the HIGH state. Then as long as the HIGH state is well separated from the other states, i.e. 
as long as $\mu_{P^{(1)}} \gg \mu_{P^{(i)}} + \sigma_{P^{(i)}}$, for $i=2,\ldots,K$, then the tail of $P(t)$ is solely determined by the HIGH state statistics and the weight of the HIGH state, 
\begin{equation}
\mathbb{P}[P(t) > d] \approx a^{(1)} \mathbb{P}[P^{(1)}(t) > d] \label{eqn:NB-mm}
\end{equation}
It therefore follows that the capacity and margin can be approximated
\begin{align}
p_r(0) & =\bar{F}^{-1}_P(\varepsilon) \approx \bar{F}^{-1}_{P^{(1)}}\left(\frac{\varepsilon}{a^{(1)}}\right) \\
m(0) & \approx \bar{F}^{-1}_{P^{(1)}}\left(\frac{\varepsilon}{a^{(1)}}\right) - \mu_P
\end{align}
For instance, consider an ON/OFF modulated Markov process with two states, 1 and 2 where the system spends $a^{(1)}$ fraction of time in state 1 on average. In the ON state, $P(t)$ is a Gaussian process with mean $\mu_{P^{(1)}}$ and variance $\sigma^2_{P^{(1)}}$; in the OFF state, $P(t)=0$, so state 1 is well separated from state 0.  Using above and Eq.~(\ref{eqn:cap-NB}),
\begin{equation}
p_r(0) \approx \mu_{P^{(1)}} + \alpha^{(1)} \sigma_{P^{(1)}}
\end{equation}
where $\alpha^{(1)} = \sqrt{2\delta^{(1)} - \ln(4\pi\delta^{(1)})}$, and $\delta^{(1)} = - \ln{(\varepsilon/a^{(1)})}$.  In other words, enough power must be provisioned to ensure that the high state has a target overdraw rate of $\varepsilon/a^{(1)}$ to ensure the overall system has a target overdraw rate of $\varepsilon$. Subtracting the overall mean, we have
\begin{equation}
m(0) \approx (\mu_{P^{(1)}} - \mu_P) + \alpha^{(1)} \sigma_{P^{(1)}}
\end{equation}
where the first term represents the power needed above the overall mean to support the ON state mean and the second term represents the power needed to suppress the ON state bursts.

These insights we draw from multi-modal distributions will be important when we study Markov modulated demand processes in Section~\ref{sec:correlated}, and again in Section~\ref{sec:evaluation} where we dissect real-world data.

%% file: SBR.tex
\section{Small Battery Regime (SBR)}
\label{sec:SBR}

This section analyzes the properties of the system when the battery is considered small relative to the demand statistics. Specifically, a ``small'' battery has sufficient stored energy to smooth out single slot, isolated, excess power demands, but is insufficient to handle excess power demands sustained over multiple time slots. To that end, we define SBR as the regime in which a battery is deployed to mitigate overdraws due to \textit{instantaneous demand spikes}. We quantify the power saved per unit of battery in the SBR. Also, for i.i.d. Gaussian processes, we determine the transition point $b_\text{tx}$ where multi-slot excess demands become significant.  The transition points for other processes are future work as the main purpose of this paper is to define, characterize, and demonstrate battery regimes.

\subsection{Ideal Bound}

In order to keep the overdraw events rare and avoid the battery nearly always being empty, there must be a positive drift to the battery state equations, i.e. $p_s$ must be at least the average power demand, $\mu_P$, and so we have $p_r(b) \geq \mu_P$ with equality only possible if $P(t)$ were deterministic. Furthermore, because any battery usage must be recharged for the battery to be of future use, the battery system cannot perform better than an equivalent system with an additional $b$ units of continuous power supply. Combining the above, we have the ideal bounds,
\begin{align}
p_r(b)&\geq \max{ \{p_r(0) - b, \space \bar{p} \}}
\label{eq:capacity-bound-ideal} \\
m(b)&\geq \max{ \{m(0) - b, \space 0\}}
\label{eq:margin-bound-ideal}
\end{align}
The ideal bound, which holds for any sample sequence $P(t)$, says that the required power and margin, cannot fall faster than linearly with battery size when measured in power charge. In terms of battery energy, $b_e = T_s b$, the required power cannot fall faster than $1/Ts$ J/W. In other words, each J of energy can save at most $1/T_s$ of provisioned power.

Perhaps surprisingly, we will show that for i.i.d. processes, these bounds are fairly tight over a significant range of battery sizes. Indeed, in the SBR, the reduction in required power is identical to the amount, $b$, of stored power charge, with every addition unit of battery storage resulting in a unit drop in required power.

\subsection{Characteristics of the SBR}
\label{sec:sbr-characteristics}

For stationary demand processes, the overdraw probability given in Eq.~(\ref{eq:overdraw-rate}) can be written as an expectation of the group's power tail probability function, 
\begin{equation}
   \mathbb{P} [overdraw] = \mathbb{E}_B\left[ \bar{F}_P(p_s + B) \right]
   \label{eqn:overdraw-rate}
\end{equation}
where the expectation is over the steady-state sample distribution of $B(t)$, denoted by the random variable $B$.
Since $B=b$ minimizes $\bar{F}_P(p_s + b)$, 
\begin{equation}
   \mathbb{P} [overdraw] \geq \bar{F}_P[p_s + b]
   \label{eq:overdraw-rate-SBR} 
\end{equation}
which also leads to the Ideal bound. In the SBR, for typical values of $\varepsilon$, the system will spend most of its time in the full battery-state and as a result the Ideal bound is tight. Therefore, in the SBR, 
\begin{align}
p_r(b) & \approx p_r(0) - b \label{eqn:capacity-SBR} \\
m(b) & \approx m(0) - b \label{eqn:margin-SBR} 
\end{align}
In the SBR, power is traded for power charge in a 1:1 ratio, e.g. $1$W of stored power charges saves exactly $1$W in required provisioned power. Intuitively, the battery is almost always full, and therefore its full capacity is available when needed; this enables the system to perform as if it had $b$ more units of constant power supply. This continues until the battery is large enough that the system transitions out of the SBR, the required power curve starts to flatten, thus there is a diminishing return on the benefits of larger batteries.  

The transition point depends on the tail statistics of the process as well as temporal correlations. For any process, define the \textit{transition battery size}, $b_\text{tx}$, as the starting point of this transition. Below $b_\text{tx}$, the system is in the SBR; above, $b_\text{tx}$, the system gradually transitions from the SBR to the LBR. Since the required power cannot drop below the mean, the margin cannot drop below $0$, and $b_\text{tx} \leq m(0)$.

It will be useful at time to express both the margin and the battery normalized to the single slot tolerable burst, $m(0)$. Specifically, define $\tilde{m}(b) = m(b)/m(0)$, and $\tilde{b} = b/m(0)$. Then the normalized version of Eq.~(\ref{eqn:margin-SBR}) becomes
\begin{align}
\tilde{m}(\tilde{b}) &= 1 - \tilde{b}, \ \text{for} \ \tilde{b} \leq \beta_{\text{tx}} \label{eqn:margin-sbr-normalized}
\end{align}
where $\beta_{\text{tx}} = b_\text{tx}/m(0)$ is the normalized SBR/LBR transition point and determines the maximum fractional margin savings in the SBR. For instance, $\beta_{\text{tx}} = .4$ means that the SBR is capable of reducing the margin requirement by 40\%, with the other 60\% falling to the LBR.

\subsection{Gaussian i.i.d. Group Power}
\label{sec:sbr-iid}

Given the linear nature of the SBR, the region is completely determined by $m(0)$ and $\beta_{\text{tx}}$.  For a Gaussian, we have already shown that $m(0) \approx \alpha \sigma_P$. So all that remains to characterize Gaussian i.i.d. demands in the SBR is to determine $\beta_{\text{tx}}$ and how it varies with the system parameters.

We have calculated Eq.~(\ref{eqn:margin-sbr-normalized}) using both simulations and exact calculations of battery state probability analytically using a discrete-time Markov-modulated process model as well as Poisson i.i.d. demands. Details of the exact Markov calculations can be found in Appendix~\ref{app:markov}.  

Fig.~\ref{fig:sbr-normalized-margin} shows the normalized margin requirement (Eq.~(\ref{eqn:margin-sbr-normalized})), computed exactly using the discrete time Markov chain model of Appendix \ref{app:markov}, for our three examples: a) Gaussian with mean 1000W and standard deviation 32W, b) Gaussian with mean 1000W and standard deviation of 79W, and c) Gaussian with mean 1000W and standard deviation of 158W. For each, the target overdraw rate is set to $\varepsilon = 10^{-3}$. Both axes are normalized to the no battery margin, $m(0)$. Also shown on the graph is the SBR line, the ideal bound of Eq.~(\ref{eq:margin-bound-ideal}). From these and numerous other examples, it is clear that for $\varepsilon = 10^{-3}$, over wide variations of means and variances, the Gaussian SBR/LBR transition occurs at roughly the midpoint, $\beta_{\text{tx}} \approx .5$, where energy storage has reduced the margin requirements by 50\%. Thus, even with small batteries, one can save a substantial amount of power without violating overdraw constraints.

The remaining 50\% reduction can be achieved in the LBR. The three curves coincide exactly: as is clear from Eq.~(\ref{eq:battery-update}) and (\ref{eqn:overdraw-rate}), the margin requirement is invariant to shifts in the mean and to scaling factors, the latter of which can be thought of as a change in units.


\begin{figure}[t]
    \centering
    \includegraphics[width=\linewidth]{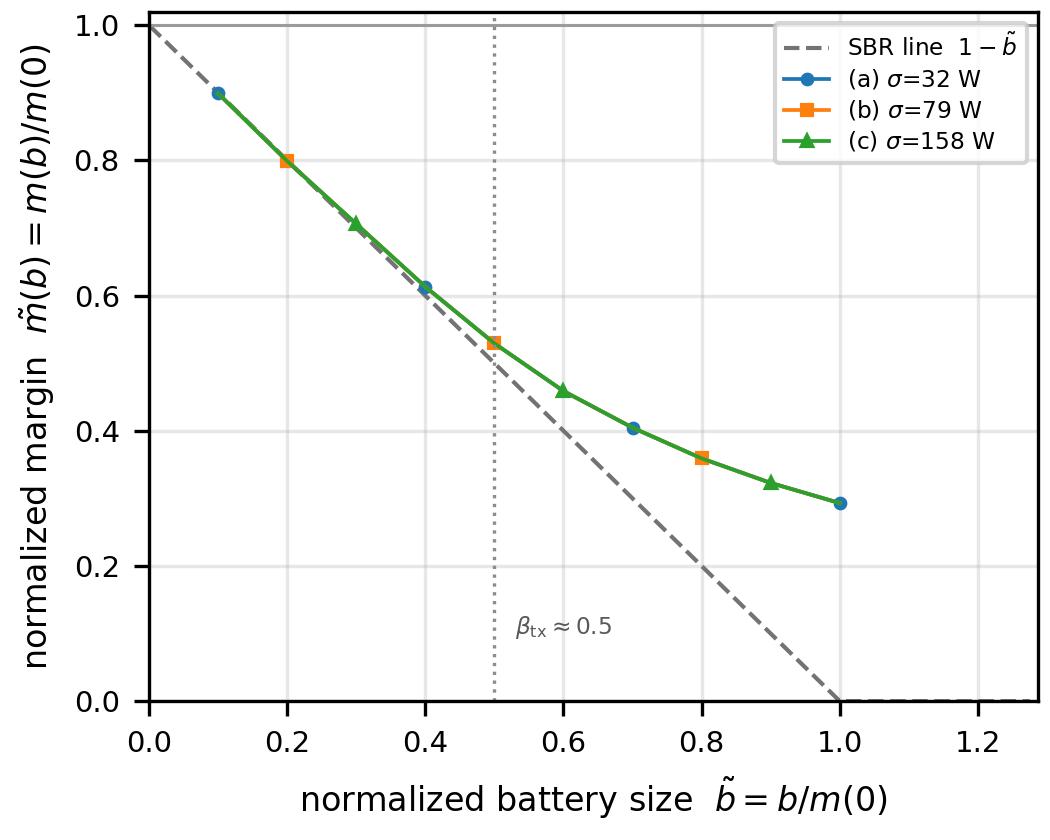}
\caption{Normalized margin $\tilde{m}(b)=m(b)/m(0)$ vs.\ normalized
battery size $\tilde{b}=b/m(0)$ for the three Gaussian i.i.d.\
examples ($\varepsilon=10^{-3}$), computed exactly via the Markov
chain model of Appendix~\ref{app:markov}. The three curves coincide
exactly: margin requirements are invariant to mean shifts and
scaling. The curves ride the SBR line $1-\tilde{b}$ (the ideal bound
of Eq.~\ref{eq:margin-bound-ideal}) until
$\beta_{\text{tx}}\approx 0.5$, then flatten as the system
transitions toward the LBR.}
    \label{fig:sbr-normalized-margin}
\end{figure}


One way to understand the transition point for i.i.d. Gaussian demands is consider what happens after an overdraw event which drains the battery. At the 50\% midpoint, the battery power charge is equal to the margin, $b = m(b) = \frac{1}{2} m(0).$ At this point, there is a $50\%$ probability that the battery will be fully recharged in the next slot, i.e. $\mathbb{P}\left(P(t) < p_s - b\right) = \mathbb{P}\left(P(t) < \mu_P \right) = \frac{1}{2}$. For larger battery sizes, this probability drops exponentially with $b$, violating the conditions of the SBR. If the demands are not i.i.d., the probability that the battery can be fully charged in the slot after an overdraw declines when the demands have positive temporal correlation, shrinking the size of the SBR.

We have repeated the above for Poisson distributions confirming that they are well modeled as a Gaussian except for small means with high positive skew. For distributions with multiple states and/or multiple modes, the SBR regime should be defined w.r.t. the highest demand state/mode. We will return to this topic when we discuss multi-state processes and compare our results to real data in Sections \ref{sec:correlated} and \ref{sec:evaluation}.

\subsection{Discussion}
\label{sec:sbr-discussion}
The two-regime structure reflects a separation of timescales. With a small battery, the battery is almost always fully charged, and overdraw occurs only when a single extreme demand spike exceeds the combined capacity of the grid and the battery. With a large battery, a single spike is easily absorbed; overdraw instead requires multiple consecutive slots of above-average demand that cumulatively drain the battery. 

Within the SBR, power savings are linear in battery size. Although Eq.~(\ref{eqn:capacity-SBR}) applies to all stationary processes, the statistics of the demand process determine the transition point out of the SBR when 2-slot overdraws become significant. 

For i.i.d. Gaussian, the SBR accounts for about $50\%$ of margin reduction. Other process are for future work, but initial results indicate:
\begin{itemize}
    \item Heavier tailed distributions require larger margins for all battery sizes, but are more efficient in that they have wider SBRs (larger $\beta_{\text{tx}}$'s).
    \item Lighter tailed distributions require less margin, but can be less efficient in that they have smaller transition points, (smaller $\beta_{\text{tx}}$'s).
    \item Short term correlations shorten the SBR (smaller $\beta_{\text{tx}}$).
\end{itemize}

%% file: large_battery.tex
\section{Large Battery Regime (LBR)}
\label{sec:eff_bw}

In this section, we will develop the concept of effective power to estimate the required power in the LBR. Effective power,  analogous to the notion of effective bandwidths~\cite{kelly1996notes} used for resource dimensioning in networking systems, is a powerful statistical characterization of the key temporal dynamics of the power demands, including variations in mean and correlations, which translates the overdraw probability constraint into a simple linear condition involving the moments of the demand distribution. 

In the LBR, batteries suppress variations in the load sustained across many time slots before an overdraw occurs.
Interestingly, in the LBR, as $b\to \infty$, the time it takes for an overdraw converges to a typical duration $T_{\text{typ}}$ with probability $1$. We derive $T_{\text{typ}}$ for a given load statistics and system parameters.



\subsection{Theory of Effective Power for Dimensioning the Demand}

The effective power of the demand process $P(t)$ is defined as 
\begin{multline}
    \label{def:eff_bw}
    \hspace{0.2in} P_\text{eff}(r) \triangleq \frac{\bar{\Lambda}_P(r)}{r} \\ = \lim_{T\to \infty} \frac{1}{rT} \log \mathbb{E}\left[ \exp\left(r\sum_{t=1}^T P(t)\right)\right].
\end{multline}
In this section, we show that for large $b$, the required power $p_r(b)$ to meet an overdraw target of $\mathbb{P}[overdraw] \leq e^{-\delta}$ can be well approximated with $P_\text{eff}\left( \frac{\delta}{b}\right)$. For a general definition and properties of effective power, see Appendix~\ref{sec:prop_eff_bw}.

Toward that end, we show that the condition
\begin{equation}
    \label{eq:eff_bw_constraint}
    p_s \geq P_\text{eff}\left(\frac{\delta}{b}\right)
\end{equation}
is sufficient to satisfy the target overdraw probability on an asymptotically tight approximation to the probability of overdraw in the LBR. This leads to the LBR required margin approximation
\begin{equation} 
m(b) \approx P_\text{eff}\left(\frac{\delta}{b}\right) - \mu_P
\label{eq:margin-LBR}
\end{equation}
Further, this approximation is asymptotically tight in $b$.

To show (\ref{eq:margin-LBR}), consider the single-layer system shown in Fig.~\ref{fig:single-layer}, where the average power demand satisfies $\mu_P < p_s$. Let the system be active since time $t=-
\infty$, so the system is in steady-state at time $t=0$. We assume that the aggregate demand is ergodic and thus the semi-invariant normalized log moment generating function exists: $\bar{\Lambda}_P(r) < \infty$ for an open region of values that includes the origin. 

Following the standard large-deviations approach for queueing and storage systems, we analyze the battery dynamics by making a simplifying assumption: we remove the finite battery boundaries in the state equation. This approximation is justified in the LBR, where the overdraw event is governed by rare threshold crossing events whose exponential decay rate is identical to that of the corresponding infinite-capacity reflected process~\cite{Ganesh2004BigQueues,Ganesh2002LDP}. Consequently, the finite battery capacity affects only lower-order (sub-exponential) terms, while the dominant exponential decay rate remains unchanged. Therefore, although the finite-capacity battery evolution given in Eq.~(\ref{eq:battery-update}) governs the exact system dynamics, the asymptotic overdraw exponent derived from the unconstrained recursion,
\[ B(t+1) = p_{s} + B(t) - P(t) \] is exponentially exact as $b\to \infty$, making it an appropriate characterization for large battery systems.

With the boundaries of the battery-state process removed, the overdraw probability originally defined in (\ref{eq:overdraw-rate}) can be further approximated as:
\begin{equation}
\label{eq:overdraw_prob_battery}
\mathbb{P}[overdraw] \approx \sup_{T>0}  \ \mathbb{P}\left(\sum_{t=-T+1}^0 P(t) > p_sT+b \right) .
\end{equation}
Note that, the value $T = T_{\text{typ}}$ which maximizes the probability in the above expression is referred to as the \textit{typical duration}for the overdraw event. We provide a formal treatment of the typical duration in Section~\ref{sec:typical_duration}. Since this approximation is asymptotically tight, moving forward, we will refer to (\ref{eq:overdraw_prob_battery}) as the overdraw probability in LBR (the counterpart in the SBR is (\ref{eq:overdraw-rate})). The tightest Chernoff upper bound on $\mathbb{P}[overdraw]$ can be written as: 
\[\mathbb{P}[overdraw] \leq \exp(-r^*b),\]
where $r^*$ is the unique positive root of equation $\bar{\Lambda}_P(r) - rp_s = 0$. Note that the Chernoff exponent is asymptotically exact, following the result:
\begin{equation}
    \label{eq:asymp_overdraw}
    \lim_{b\to \infty} \frac{1}{b}\log \mathbb{P}[overdraw] = -r^*,
\end{equation}
which is a consequence of a more general theorem proved in~\cite{deveciana1995effective}. 

Following the Chernoff upper bound, a sufficient condition to meet a desired overdraw probability, $\mathbb{P}[overdraw] \leq e^{-\delta}$ can be stated as:
\begin{equation}
\label{eq:limit_constraint}
\mathbb{P}[overdraw] \leq \exp(-r^*b) \leq \exp(-\delta) ~ \Rightarrow ~ r^*>\frac{\delta}{b}.
\end{equation}
Since $\bar{\Lambda}_P(r) - rp_s$ is convex with two roots: $r=0$ and $r=r^*$, for all values in $(0,r^*),\ \bar{\Lambda}_P(r) - rp_s<0$. Thus, (\ref{eq:limit_constraint}) implies:
\[ \bar{\Lambda}_P\left(\frac{\delta}{b}\right) - \frac{\delta}{b} p_s \leq 0, \]
which directly leads to the condition stated in (\ref{eq:eff_bw_constraint}), and thus
the amount of supply power necessary to meet the overdraw constraint $e^{-\delta}$ is no more than $P_\text{eff}\left(\frac{\delta}{b} \right)$. Further, since this upper bound is asymptotically tight, we have 
\begin{equation}
p_r(b) \approx P_\text{eff}\left(\frac{\delta}{b} \right) 
\end{equation}
which leads to the LBR margin approximation (\ref{eq:margin-LBR}).

Due to the additive nature of effective power, the above arguments generalize to the situation with multiple servers/racks multiplexed with statistically independent load characteristics:
\begin{equation}
    \label{eq:eff_bw_constraint_mux}
    \sum_{k=1}^n P_{\text{eff},k}\left(\frac{\delta}{b}\right) \leq p_s.
\end{equation}
The condition expressed in (\ref{eq:eff_bw_constraint_mux}) illustrates that the required margin is proportional to the combined burst of the cumulative load.  Typically, as the servers/racks are multiplexed, the power supply is also scaled accordingly. We will address this topic in Section \ref{sec:stat-mux}.

\subsection{Gaussian Demands and the Second-Order Approximation} 

Closed form expressions for $P_\text{eff}$ are possible for some demand processes. For instance, for i.i.d. Gaussian processes, $P_\text{eff} \approx \mu_P+ \frac{1}{2} \frac{\sigma_\text{P}^2}{b} \delta$. For others, numerical calculations are straightforward but do not provide closed form expressions which can provide physical insight. 

To that end, we derive a second order approximation to $P_\text{eff}$ (see Appendix~\ref{sec:prop_eff_bw} for the full expression), 
\begin{equation}
    \label{eq:eff_bw_constraint_mux_second_order}
    P_\text{eff}\left(\frac{\delta}{b}\right) \approx \mu_P + \frac{1}{2} \frac{\sigma_\text{eff}^2}{b} \delta,
\end{equation}
where 
\begin{align} 
\nonumber 
\sigma_\text{eff}^2 &\triangleq \lim_{T\to \infty}\frac{1}{T} \text{var}\left( \sum_{t=1}^T P(t) \right) \\
&=\sigma_{P}^2+\lim_{T\to \infty} \frac{1}{T}\sum_{t\neq \tau} \sum_{\tau=1}^T \text{cov}(P(t),P(\tau)) .
\label{eq:on-off-cov-1}
\end{align}
is the \textit{effective variance} of the process. The effective variance consists of two parts: the variance of the demand and a term which encapsulates temporal correlations. For i.i.d. demands, $\sigma_\text{eff} = \sigma_P$.  With temporally positively correlated demands, $\sigma_\text{eff}^2>\sigma_P^2$.

The accuracy of the second-order approximation depends on both the skew of the underlying single slot probability demand $P(t)$ as well as the size of the battery compared to how fast the temporal correlations decay. For Gaussian demands (potentially correlated), the approximation is exact. For general demand statistics, it becomes accurate for $b$ large enough to average out the demand fluctuations due to temporal correlations. We will discuss this further in Section~\ref{sec:correlated}.

The second-order approximation clarifies how margin requirements scale with system parameters. With no probabilistic constraint ($\delta=0$) , it suffices to set the available power identical to the mean load $p_r(b) =\mu_P$. For any probabilistic constraint, i.e., $\delta > 0$, the \textit{minimum} amount of margin that we need to set aside to meet that constraint is approximately
\begin{equation} 
m(b) \approx \frac{1}{2} \frac{\sigma_\text{eff}^2}{b} \delta.
\label{eq:margin_second_order}
\end{equation}
For any given battery size, the margin increases with the effective variance of the power demands and falls as $\frac{1}{b}$ rather than the linear decay in the SBR.  Also, notice that the margin requirement is not that sensitive to the overdraw rate since $\delta = -\log(\varepsilon)$.

\begin{figure}[t]
    \centering
    \includegraphics[width=\linewidth]{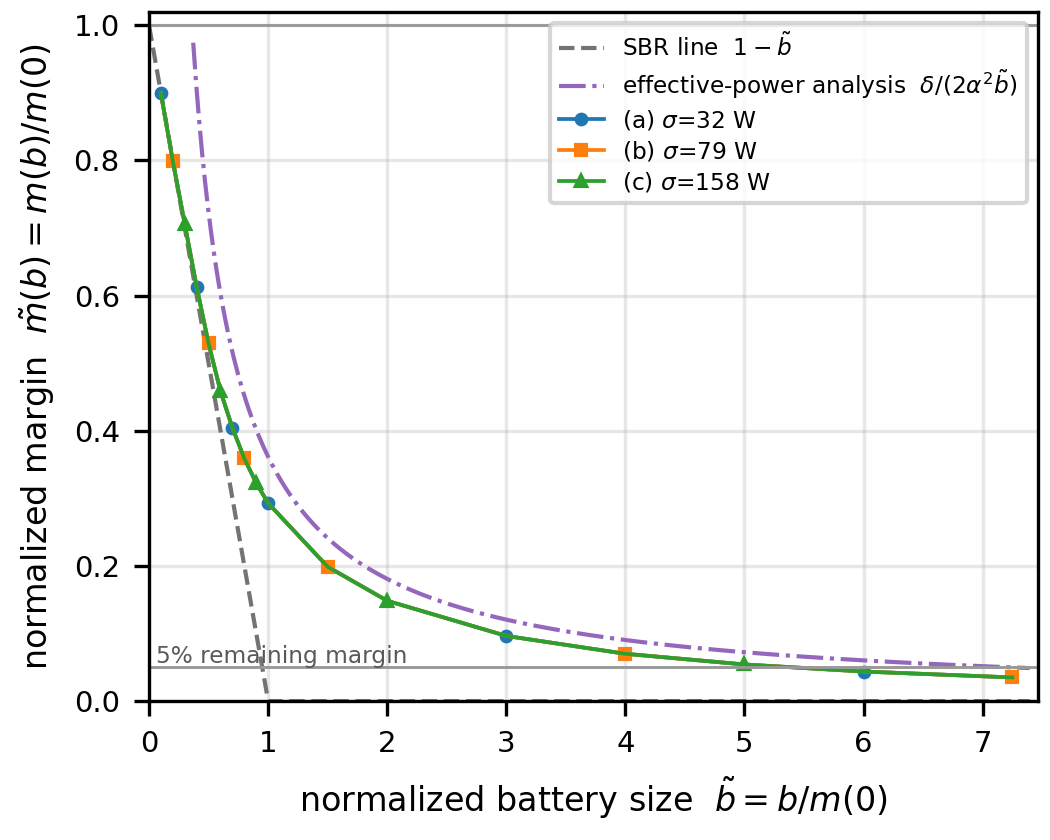}
    \caption{The complete normalized margin curves for the three
    Gaussian i.i.d.\ examples of Fig.~\ref{fig:sbr-normalized-margin}
    ($\varepsilon=10^{-3}$, computed exactly via the Markov chain model
    of Appendix~\ref{app:markov}), continued into the LBR. Beyond the
    SBR/LBR transition the exact margin decays as $1/\tilde{b}$; the
    second-order effective-power approximation
    $\delta/(2\alpha^2\tilde{b})$ (dash-dotted) captures this decay and
    gives a conservative, slightly over-provisioning estimate of the
    required margin. The exact curves reach 5\% remaining margin near
    $\tilde{b}\approx 5.3$.}
    \label{fig:sbr-normalized-margin-full}
\end{figure}

Fig.~\ref{fig:sbr-normalized-margin-full} extends the three Gaussian examples of Fig.~\ref{fig:sbr-normalized-margin} into the LBR and compares the exact margin with the second-order effective-power approximation of Eq.~(\ref{eq:margin_second_order}). Although the approximation is derived asymptotically (as $b\to\infty$), it becomes
useful at practical battery sizes where it captures the $\frac{1}{b}$ decay and provides a conservative estimate of required power. 

\subsection{Typical Duration for an Overdraw Event}
\label{sec:typical_duration}

The overdraw event given in (\ref{eq:overdraw_prob_battery}), 
\[ \sum_{t=-T+1}^0 P(t) > p_sT+b ,\]
potentially occurs in a variety of ways depending on the value of $T$. It may occur for $T=1$ with an instantaneous burst in the load, or it may take arbitrarily long with the load exceeding the supply just slightly, sustained over $T\gg 1$. In the LBR, neither extreme turns out to be the way an overdraw occurs. Instead, there is a typical duration $1<T_\text{typ}(b)<\infty$ for the overdraw event to occur, 
\begin{equation}
\label{eq:typical_duration}
T_\text{typ}(b) \triangleq \underset{T>0}{\arg\max}\ \mathbb{P}\left(\sum_{t=-T+1}^0 P(t) > p_sT+b \right)
\end{equation}
This follows from Laplace's principle, which states that when an unlikely event occurs, there is a typical way for it to occur, whose probability dominates over all other possibilities (see~\cite{varadhan1966asymptotic,dembo1998} for formalization). 

 $T_\text{typ}(b)$ is derived in Appendix~\ref{sec:T_typ},
\begin{equation}
    T_\text{typ}(b) = \frac{b}{\bar\Lambda_P'(r^*)-p_s}.
    \label{eq:typical_time_general}
\end{equation}
where $r^*$ is defined Appendix~\ref{sec:T_typ}. The typical overdraw duration is proportional to the battery capacity and thus larger and larger batteries take longer to drain for a fixed provisioned power $p_s$. 

For example, for an i.i.d. Gaussian power demand with a mean $\mu_P$ and variance $\sigma_P^2$,  $r^*=\frac{2(p_s-\mu_P)}{\sigma_P^2}$ and the typical overdraw
\begin{equation}
   T_\text{typ}(b)=\frac{b}{p_s-\mu_P}
   \label{eq:Gaussian_typical_duration}
\end{equation}
turns out to be only a function of the mean demand, and is independent of the higher order moments, including the variance. Similarly for i.i.d. Poisson demand with rate $\lambda=\mathbb{E}[P(t)]$, we can obtain the exact same form: $T_\text{typ}(b) \rightarrow \frac{T}{p_s-\lambda}$ as $\lambda \to p_s $. We leave this as an exercise to the reader. Thus, increased supplied power reduces $T_{\text{typ}}$ as it becomes less likely that the power will exceed demand over any period of time. 

Furthermore, when the supply is set identical to the required power, $p_s = p_r(b)$, so that the overdraw constraint is met exactly, it follows from Eqns.~(\ref{eq:Gaussian_typical_duration}) and (\ref{eq:margin_second_order}), that $T_\text{typ}(b) = O(b^2)$. 

\subsection{Decay Rate of Overdraw Probability}

We can draw further insights by comparing the overdraw rates with and without battery. We defined rate function $I_P(p_s)$ for NBR in Eq.~\ref{eq:rate-function}, which dictates the rate of decay in the overdraw probability $\mathbb{P}[overdraw] \sim \exp(-I_P(p_s))$, as provided in (\ref{eq:chernoff_no_battery}). In Appendix~\ref{sec:T_typ}, we develop an analogous rate function $I_B(p_s)$ for LBR and show in (\ref{eq:rate_battery}) that $ \mathbb{P}[overdraw] \sim \exp (- T_\text{typ}(p_s) I_B(p_s) )$. Geometrically, the two rate functions are also illustrated in Fig.~\ref{fig:log_mmt} in Appendix~\ref{sec:T_typ}. Both rate functions $I_P$ and $I_B$ increase with the margin, i.e. the higher the supplied power over the mean the faster the decay rate in overdraw probability. However, comparing the exponents of the overdraw probabilities, we observe further insights into the how the overdraw events occur in these different regimes:
\begin{enumerate}
        \item $I_P \ll I_B$. The difference in the rate function underscores the additional amount of burst required in the overdraw event. Without the battery, it suffices that the cumulative load instantaneously exceeds the supply $p_s$. On the other hand, an overdraw can only occur with a battery, if the load exceeds the supply $p_s$ by a positive amount. 
        \item $1 \ll T_\text{typ}(b)$. The difference in the typical duration for the overdraw event. Without a battery, the overdraw event is instantaneous, i.e., the time-scale for an overdraw event is $1$. However, with a battery, the typical time scale is $T_\text{typ}(b)$. Thus, in this regime, the load burst should not only exceed the supply by a positive amount, but also be sustained for a long period of time, identical to $T_\text{typ}(b)$, which scales with $b$.
\end{enumerate}

%% file: stat_mux.tex
\section{Statistical Multiplexing of Energy Demand}
\label{sec:stat-mux}

A key architectural question in designing the system in Fig. \ref{fig:layered-arch} is determining the size of an energy group. It is well known that increasing the number of servers $n$ sharing a power supply increases efficiency by reducing the required margin per server, $q_n(b) \triangleq  m(b)/n$. This section investigates how efficiency scales with battery and group size. We will consider both finite and asymptotic results. 

Shared energy systems benefit from two forms of statistical multiplexing:
\begin{itemize}
    \item Spatial multiplexing: the pooling of multiple power demands into one aggregate demand served by one power supply.
    \item Temporal multiplexing: the use of a battery to smooth out temporary power fluctuations.
\end{itemize}
Both forms of multiplexing reduce the required margin per server. Ignoring potential negative correlations, the efficiency gains of spatial multiplexing are greatest when server power demands are independent, whereas the efficiency gains of temporal multiplexing are greatest when time slots are independent. Both types exhibit diminishing return behavior, and thus we would expect the efficiency gains of the combination of them to be less than cumulative. In fact, as will be shown, increasing one form of multiplexing may have a much smaller effect than anticipated by the naive combination of the two.

To quantify the efficiency gains, Section \ref{sec:stat-mux-2ndOrder} derives $q_n(b)$ for groups with demands well modeled by the second order approximations (\textit{2nd-order demands}). The benefits of spatial and temporal multiplexing are separated and general qualitative scaling results are discussed.

Then in Section \ref{sec:stat-mux-iid}, we focus the discussion on such groups where the server demands are independent and temporally uncorrelated gaussian processes. For those, we consider 3 scaling scenarios:
\begin{itemize}
    \item \textbf{Scenario 1:} Fixed $b$, increasing $n$.
    \item \textbf{Scenario 2:} $b$ and $n$ scale together, with $b \propto \sqrt{n}$
    \item \textbf{Scenario 3:} $b$ and $n$ scale together, with $b \propto n$
\end{itemize}
and quantify  $q_n(b)$ for each. In Scenarios 1 and 2, the deployed battery per server, $b/n$, decreases with increased group size, where it is constant in Scenario 3. We show that key to understanding these different scenarios is how fast the battery capacity grows relative to the SBR/LBR transition point, $b_{\text{tx}}$. Asymptotically, Scenario 2 is the natural energy capacity growth which balances temporal and spatial multiplexing. In Scenario 1, battery capacity grows too slowly and spatial multiplexing asymptotically dominates whereas in Scenario 3, the battery capacity grows fast enough such that both forms of multiplexing meaningfully contribute to efficiency asymptotically.

Section \ref{sec:stat-mux-iid} summarizes the results.

\subsection{The Combined Effects of Spatial and Temporal Multiplexing}
\label{sec:stat-mux-2ndOrder}
In general, given statistical characterization of the group power demand, $P(t)$, the margin/server $q_n(b)$ can estimated using results presented in the last sections, specifically Eqns. (\ref{eqn:margin-SBR})  for SBR and (\ref{eq:margin-LBR}) for LBR. The margin/server will depend on the servers' demand statistics, particularly its tail, and their inter- and intra-correlations, all of which may change with group size. Complicating matters, depending on how fast energy storage scales with the number of servers, the system may move from one regime to another, e.g. from LBR to SBR or vice-versa, and the transition region between them may shrink or grow. 

In the next subsection we will analyze $q_n(b)$ for Gaussian i.i.d. processes. Here we consider a more general case but only consider how the margin asymptotically scales with the system parameters. To that end, we make the simplifying assumption that for large enough group size the group margin requirement is well approximated by the second order approximations. In these cases, the margin/server can be estimated by dividing (\ref{eqn:margin-SBR}) and (\ref{eq:margin_second_order}) by $n$.  After some manipulation, and using $q_n(0) \approx \sqrt{2 \delta} \sigma_P$ (Eqn. \ref{eqn:NB-margin-2ndOrder}), the margin/server in each regime can be approximated
\begin{align}
\text{SBR:}\ q_n(b) & \approx  q_n(0) \left( 1 - \frac{b/n}{q_n(0)}\right)  \\
\text{LBR:}\ q_n(b) & \approx  q_n(0) \left(\frac{\phi_n }{4} \frac{q_n(0)}{b/n}\right)
\label{eqn:margin/server}
\end{align}
where $\phi_n \triangleq \sigma_{\text{eff}}^2/\sigma_P^2 \geq 1$ is a measure of the temporal correlation of the group demand process with $\phi_n = 1$ for uncorrelated demands. In each regime, the term $q_n(0)$ is the portion of the margin solely attributable to spatial multiplexing and the second terms are the efficiency gains achieved by adding energy storage.

The growth of $q_n(0)$ is solely determined by the single slot joint statistics of the server demands, invariant to any temporal correlations. For demand processes considered in this paper, $q_n(0)$ will scale between  $O(1/\sqrt{n})$ for statistically independent demands and $O(1)$ for perfectly correlated demands. 

On the other hand, the efficiency gains of temporal multiplexing also depends on the amount of energy storage, temporal correlations in the LBR, and the SBR/LBR transition, $b_{\text{tx}}$, all of which change with group size. For our purposes here, we ignore the transition region between the SBR and LBR and approximate the transition as the point where the typical overdraw rate $T_{\text{typ}} \geq 1$. From Eqn. (\ref{eq:Gaussian_typical_duration}) and using the LBR approximation $m(b) \approx \frac{\delta \sigma_{\text{eff}}^2}{2b}$, 
\begin{equation}
\frac{b_{\text{tx}}}{n} \approx \frac{q_n(0)}{2} \sqrt{\phi_n}
\end{equation}
For demand processes considered in this paper, $\phi_n = O(1)$ and thus asymptotically $q_n(0)$ defines the \textit{natural growth rate of the battery per server}, i.e. the required growth rate of the battery/server to keep the amount of temporal multiplexing gain constant.  This rate varies between $O(1/\sqrt{n})$ for statistically independent demands and $O(1)$ for perfectly correlated demands. 

Thus, asymptotically,
\begin{itemize}
    \item if $b/n$ grows slower than the natural rate, e.g. if the battery size is fixed, then increasing group size improves efficiency through spatial statistical multiplexing, but dilutes the effects of temporal multiplexing by reducing the relative battery size. Above some $n$, the system will be in the SBR and the benefits of energy storage will be diluted away until the system behaves approximately like a no battery system. Thus asymptotically $q_n(b)$ scales with $q_n(0)$, i.e. between $O(1/\sqrt{n})$ and $O(1)$ for independent to perfectly correlated demands.
    \item If $b/n$ grows at the natural rate, the system will stay in the SBR or LBR, as appropriate. Like above, $q_n(b)$ scales between $O(1/\sqrt{n})$ and $O(1)$ for independent to perfectly correlated demands, with temporal multiplexing providing a constant multiplicative efficiency improvement. 
    \item If $b/n$ grows faster than the natural rate, then above some $n$, the system will be in the LBR and $q_n(b)$  scales with $q^2_n(0)/(b/n)$. For example, if the demands are independent and the amount of energy storage/server is kept constant, then $q_n(b) = O(\frac{1}{n})$.
\end{itemize}

\subsection{Independent Gaussian i.i.d. Demands}
\label{sec:stat-mux-iid}

The above analysis can be refined for independent gaussian temporally uncorrelated demands. In this case, $\phi_n = 1$ and $\sigma_P^2 = n \sigma_X^2$. Also since both $X_i(t)$ and $P(t)$ are gaussian, $q_n(0) = q_1(0)/\sqrt{n}$ where $q_1(0) \approx \alpha \sigma_X$ and the transition between the SBR/LBR occurs at roughly $50\%$ margin reduction. Using these in Eqs., (\ref{eqn:margin-SBR}) and (\ref{eq:margin_second_order}) and dividing by $n$, we have
\begin{align}
\text{SBR:}\ q_n(b) &\approx  \frac{q_1(0)}{\sqrt{n}} \left( 1 - \frac{b}{q_1(0) \sqrt{n}}\right) \\
\text{LBR:}\ q_n(b) &\approx  \frac{q_1(0)}{\sqrt{n}} \left(\frac{\delta}{2\alpha^2}\frac{q_1(0) \sqrt{n}}{b}\right)  \\
\text{SBR/LBR transition:}\ b_{\text{tx},n} &\approx .5  \space q_1(0) \sqrt{n} \label{eqn:margin/server-ind}
\end{align}
and the natural growth rate of the total battery requirement is $O(\sqrt{n})$. The constant $\frac{\delta}{2 \alpha^2}$ varies between .45 and .3 for target overflow rates between $10^{-2}$ and $10^{-6}$.

Now consider the 3 growth scenarios: 1) $b$ constant, 2) $b \propto \sqrt{n}$, and 3) $b \propto n$. 

\subsubsection{Scenario 1: Fixed $b$, increasing $n$}

In Scenario 1, the number of servers increases without adding additional storage. In this case, the energy stored/server decreases as $1/n$ and the system enters the SBR when the temporal correlation term is about $50\%$, i.e. when $n \geq \left(\frac{2b}{q_1(0)}\right)^2 = \frac{4 b^2}{\alpha^2 \sigma_X^2}$.  As $n$ increases, the relative contribution of temporal multiplexing becomes increasingly small, eventually becoming insignificant. Asymptotically, the system will behave like one without any energy storage and any efficiency gains are due to spatial multiplexing with $q_n(b) \xrightarrow{n \to \infty} q_1(0)/\sqrt{n}$.

\begin{figure}
    \centering
    \includegraphics[width=1\linewidth]{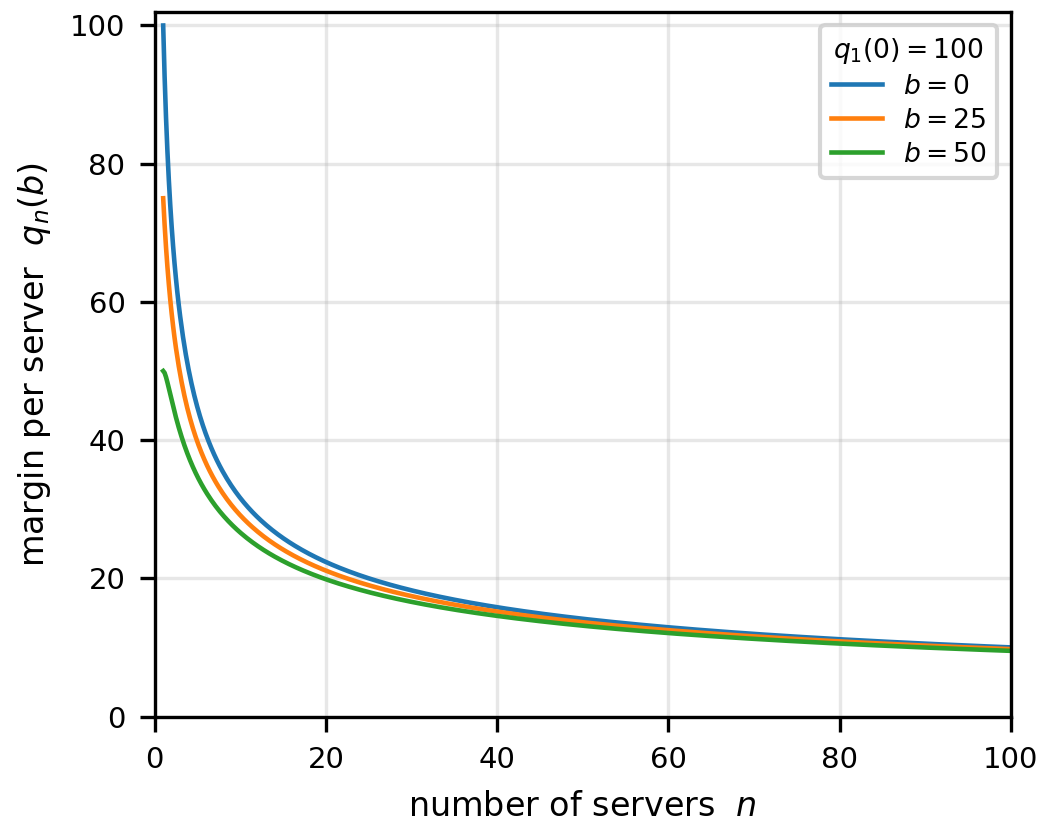}
    \caption{The combined effects of multiplexing with fixed energy storage (Scenario 1)}
    \label{fig:stat-mux-fixed-B}
\end{figure}

However, for finite $n$, both temporal and spatial multiplexing can contribute meaningfully to increased efficiency. Fig. \ref{fig:stat-mux-fixed-B} shows $q_n(b)$ as a function of $n$ with $q_1(0)=100$ and $b=0, 25, 50.$ The top curve shows the benefits of pure spatial multiplexing with $q_n(0)$ falling as $1/\sqrt{n}$. For very small $n$, the margin/server can be significantly reduced by adding battery capacity. However, as $n$ grows, the increase in spatial multiplexing is offset somewhat by a decrease in temporal multiplexing, slowing the margin/server reduction until all curves catch up to each other as the effect of the battery is diluted away. 

\subsubsection{Scenario 2: $b \propto \sqrt{n}$}

In Scenario 2, battery capacity is added with additional servers at a rate of $b = \xi\sqrt{n}$ for some constant $\xi$. In this case, the battery grows at the same rate as the SBR/LBR transition and the system will stay in the SBR or LBR, depending on the value of $\xi$. Thus, the contribution due to temporal multiplexing is fixed, and asymptotically the margin/server fall as $1/\sqrt{n}$  as in Scenario 1. However, unlike Scenario 1, the benefits of temporal multiplexing at not diluted away, but rather provides a multiplicative reduction to the decay rate for any $n$. Specifically, for $b= \xi\sqrt{n}$ for some constant $\xi$, we have
\begin{align}
\text{SBR:}\ q_n(b) &\approx  \frac{q_1(0)}{\sqrt{n}} \left( 1 - \frac{\xi}{q_1(0)}\right) \\
\text{LBR:}\ q_n(b) &\approx  \frac{q_1(0)}{\sqrt{n}} \left(\frac{\delta}{2\alpha^2}\frac{q_1(0)}{\xi}\right)
\end{align}
with no transition between the two regimes as $n$ and $b$ grow. Thus, asymptotically $q_n(b) \xrightarrow{n \to \infty} c q_1(0)/\sqrt{n}$ where $c$ is either the SBR or LBR temporal multiplexing term above.

\subsubsection{Scenario 3: $b \propto n$}

In Scenario 3, battery capacity is added with additional servers at a rate of $b = \gamma n$ for some constant $\gamma$.  Here 
\begin{align}
\text{SBR:}\ q_n(b) &\approx  \frac{q_1(0)}{\sqrt{n}} \left( 1 - \frac{\gamma \sqrt{n}}{q_1(0)} \right) \\
\text{LBR:}\ q_n(b) &\approx  \frac{q_1(0)}{\sqrt{n}} \left(\frac{\delta}{2\alpha^2}\frac{q_1(0)}{\gamma\sqrt{n}} \right)  \\
\text{SBR/LBR transition: }\ & n \approx  \frac{\alpha\sigma_x^2}{2 \gamma}  \label{eqn:margin/server-linear b}
\end{align}
Thus, for large enough $n$ the system is the LBR and $q_n(b) \xrightarrow{n \to \infty} \frac{1}{n}$, with spatial and temporal multiplexing equally contributing a $1/\sqrt{n}$ factor to efficiency.

\subsection{Discussion}
\label{sec:stat-mux-discussion}

For systems to asymptotically benefit from temporal multiplexing, the battery must grow commensurate with the no battery required margin. Except for perfectly correlated demands, increasing the group size reduces the required battery/server for a given amount of temporal multiplexing. For uncorrelated demands, the battery/server can be reduced as $O(1/\sqrt{n})$ and still maintain the same efficiency gains. However, for highly correlated demands, the total battery storage must grow about linearly with the group size.

%% file: correlated.tex
\section{Multi-scale Temporally Correlated Power Demands}
\label{sec:correlated}

To capture the stochastic structure of data center loads at multiple time scales, we model the power demand as a Markov-modulated processes (MMP). These processes provide an explicit multi-scale representation: the underlying Markov chain governs slow transitions between operating regimes (e.g., low-load, bursty, or sustained high-load states), while the i.i.d. arrivals within each state capture fast fluctuations. Second, these models naturally embed the persistence of temporal correlations in high-demand periods. We demonstrate the fit of the model to real data in the next section.

The section is organized as follows. First, we formally present the multi-state model. Next, we derive approximations to the no-battery, small battery, and large battery capacities. Finally, we apply these results to a simple ON/OFF process.  Within that analysis, we also provide intuitive approximations to each region of battery state. 

\subsection{Multi-state Model}

The group power $P(t)$ is modeled as a stochastic process modulated by a $K$-state Markov chain where the probability of a transition from state $i$ to state $j$ is denoted ${\cal P}_{ij}$. Transitions from one state to another occur at the end of time slot, so the state at time $t,\ J(t)$,  solely determines the statistics of $P(t)$.  Further, we assume that the Markov chain is irreducible and positive recurrent and therefore has stationary state probabilities  $\pi_k, k = 1,\ldots,K$.  

When the chain enters state $k$, it will stay for a geometrically distributed number of slots with \textit{average} \textit{dwell time} $L_k = 1/(1-\mathcal{P}_{kk})$. As it pertains to the battery dynamics, systems with large dwell times will behave differently than those with small times, even if they have the same stationary distributions. Long dwell times and widely separated states model long term cyclo-stationary power demands, whereas short dwell times between relatively close states model intra-period correlations.

While in state $k$, the group's power demand is described as $P_k(t)$, which we assume to be i.i.d. with mean $\mu_k$, standard deviation $\sigma_k$, moment generating function $g_k(r)$, and with tail and quantile functions $\bar{F_k}(d)$ and $\bar{F}^{-1}_k(\varepsilon)$, respectively. Without loss of generality, order the states so means are decreasing. State $1$ is called the HIGH state and state $K$, the LOW state. Note the average power demand is $u_P = \sum_{k=1}^{K} \pi_k u_k$.

\subsection{No-battery Margin Requirements}

Following the approach in Section \ref{sec:baseline}, the no battery margin can be calculated from the quantile function (Eq. \ref{eqn:margin-NB}),
\begin{align}
m(0) & \approx \bar{F}^{-1}_1\left(\varepsilon \right) - \mu_P
\end{align}
Unlike the single state case, approximating/bounding the multi-state $m(0)$ using a second order approximation to Chernoff bound will not produce accurate results because of the highly skewed multi-modal form of the pdf of $P(t)$.

However, in the case where the HIGH state is well separated from the rest of the states, i.e., $\mu_1 - \mu_k \gg \sigma_k$ and the system spends an amount of time in each state such that $\pi_k \gg \varepsilon > 0$, for $k=2,\ldots,K$, then the probability distribution of  $P(t)$ is multi-modal, and following the analysis in Section \ref{sec:multi-mode-NB}, the margin can be approximated as
\begin{align}
m(0) & \approx \bar{F}^{-1}_1\left(\frac{\varepsilon}{\pi_1} \right) - \mu_P \label{eqn:margin-nb-high}
\end{align}
For instance, if the target overflow rate is $\varepsilon=10^{-3}$, and if the system spends half its time in the high state, then there must be sufficient power to guaranty an overflow probability of no more than $0.002$ in the high. Note that, $m(0)$ depends on the amount of time spent in the HIGH state, but not on how it spends it time there, i.e. its dwell time.

Now, using a second order approximation to the Chernoff bound on the HIGH state only,
\begin{equation}
m(0) \approx (\mu_1 - \mu_P) + \alpha_1 \sigma_1  \label{eqn:margin-NB-MMP-gaussian}
\end{equation}
where $\delta_1 = - \log{(\varepsilon/\pi_1})$ and $\alpha_1 = \sqrt{2\delta_1}$ (or $\sqrt{2\delta_1 - \ln(4\pi\delta_1)}$ for Gaussian).  

Expressed this way, the no-battery margin has two components: one due to the HIGH mean being larger than the average, and one due to HIGH state power fluctuations above its mean. The fraction of time spent in the HIGH state, $\pi_1$, has only a small effect on the first term and none on the second term. We will see below that smaller batteries suppress HIGH fluctuations whereas much larger batteries are needed to reduce the latter.

\subsection{SBR Margin Requirements}
 
For small enough $b$, we expect that all systems exhibit some small battery region where $p_r(b) \approx p_r(0) - b $, with the transition out of SBR dependent on demand statistics.

In cases where the HIGH state \textit{dominates}, i.e. is widely separated from the remaining states and the system dwells in the HIGH state for a reasonable period of time, the SBR region is determined mainly by the HIGH state, with the SBR/LBR transition occuring at approximately  
\begin{equation}
b_\text{tx} = \beta_1 (p_r(0) - \mu_1)
\end{equation}
where $\beta_1$ is the fractional transition point for the HIGH state and determined by the statistics of the HIGH state. For instance, if the HIGH state is Gaussian, 
\begin{align}
p_r(b) &\approx \mu_1 + \alpha_1 \sigma_1 - b \label{eqn:cap-sbr-mmp} \\
m(b) &\approx (\mu_1 - \mu_P) + (\alpha_1 \sigma_1 - b)
\end{align}
and the system transition out of the SBR at the \textit{first midpoint} where the capacity is halfway between the no-battery capacity and HIGH mean, $b_\text{tx} = .5 \alpha_1 \sigma_1$.  At this point,
\begin{align}
p_r(\beta_\text{tx}) &= \mu_1 + .5 \alpha_1 \sigma_1 \\
m(\beta_\text{tx}) &= (\mu_1 - \mu_P) + .5 \alpha_1 \sigma_1
\end{align}
Thus, in the SBR, the battery can some of the HIGH state bursts.

\subsection{LBR Margin Requirements}

The LBR region starts after the SBR when the overflow times start to exceed a single slot. In this region, the required power to meet the overflow requirement is approximately the effective power. Since the process is ergodic but not i.i.d.,
\begin{equation}
p_r(b) \approx P_{\text{eff}} \left(\frac{\delta}{b}\right) = \frac{b}{\delta} \bar{\Lambda}_P\left(\frac{\delta}{b}\right)
\end{equation}
The effective power can be computed by taking the log of the Perron-Frobenius eigenvalue, $\rho(\cdot)$, of the tilted transition matrix with entries $K_{ij}(r) = {\cal P}_{ij} \cdot g^{(i)}(r)$, 
\begin{equation}
P_{\text{eff}}(r) =\frac{1}{r} \log \rho(K(r))  \label{eq:MMPP_log_mmt}
\end{equation}
which comes from the definition of $P_{\text{eff}}(r)$ and the fact that $\bar{\Lambda}_P(r)=\log \rho(K(r))$\cite{tse1995multiplexing}.

Even for Gaussian i.i.d. states, the solution to Eq.~(\ref{eq:MMPP_log_mmt}) is complex and will display many LBR subregions with differing characteristics depending upon the statistical properties of the states, their separation, and their transition dynamics. In general, calculation of the effective power requires numerical computation of Eq. (\ref{eq:MMPP_log_mmt}). 

When the states are widely separated with long enough dwell times, we can gain physical insight by approximating Eq.~(\ref{eq:MMPP_log_mmt}) at the start and end of the LBR using this paper's previously developed tools. For instance, when the high state dominates, we can approximate $p_r(b)$ at the start of the LBR using the theory of effective power in that state alone. If the demand process is not too skewed, then the margin at the start of the LBR will fall as $1/b$ asymptotically to the high state mean until the system transitions to the next LBR subregion. Similarly, when the battery $b$ is large enough to average out temporal correlations, the margin in the deep LBR can be approximated as $m(b) \approx \frac{\delta \sigma_{\text{eff}}^2}{2b}$, where $\sigma_\text{eff}^2$ was defined in Eq.~(\ref{eq:on-off-cov-1}). However, these two approximations are insufficient to describe the full LBR even for a simple ON/OFF process, as is examined next.

\subsection{ON/OFF Example} \label{sec:on-off}

In this section we consider a simple 2 state ON/OFF process where the ON state mean is $\gg 0$.  In this case, 
\[ K(r) =  \begin{bmatrix} 
\left(1 - \frac{1}{L_1}\right) g_1(r) & \frac{1}{L_1} \space g_1(r) \\ 
\frac{1}{L_2} & 1 - \frac{1}{L_2} \end{bmatrix}  \]
where state 1 is the ON state with dwell time $L_1$, and state 2 is the OFF state with dwell time $L_2$. When the ON demand process is i.i.d. Gaussian, $g_{1}(r) = \exp\left( \mu_{1} r + \frac{1}{2}\sigma_{1}^2 r^2 \right)$. The mean and variance are easily calculated 
\begin{align}
\mu_P & = \pi_1 \mu_1 \\
\sigma^2_P & = \pi_1 \sigma^2_1 + \pi_1(1-\pi_1) u_1^2
\end{align}
Note that both are functions only of the time spent in the ON state, not its dwell time.  Since the required power $p_r(b)$ will depend on the dwell times, the second-order approximation to $p_r(b)$ presented earlier is clearly insufficient. However, similar to the single-state case, the exact power requirement $p_r(b)$ can be numerically calculated by modeling the system as a 2-dimensional Markov chain with the state space $(B(t),P(t))$.  

\begin{figure}[t]
\centering
\includegraphics[width=\columnwidth]{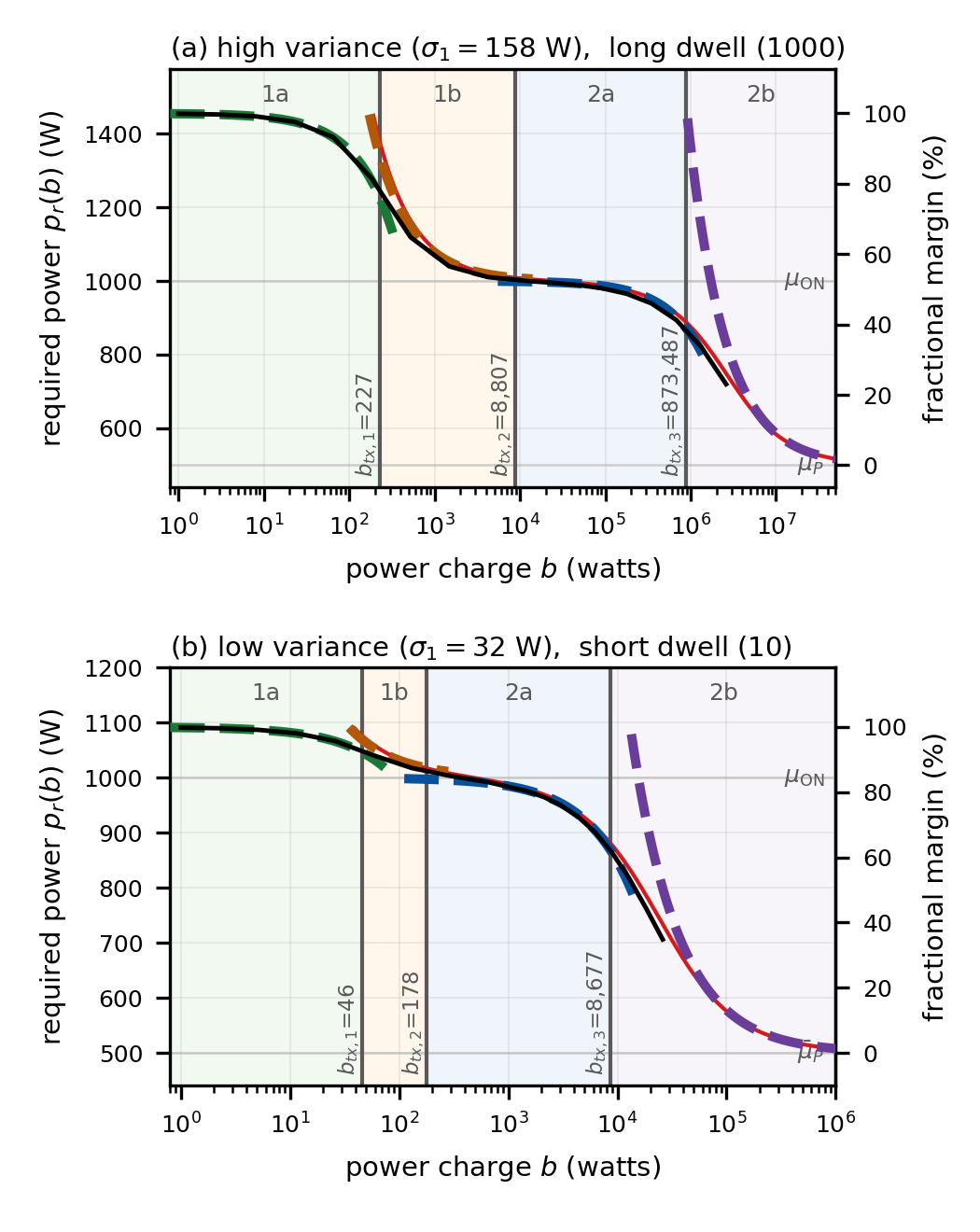}
\caption{Power required to meet overload target
$\varepsilon=10^{-3}$, versus battery power charge $b$ for an ON/OFF
Markov-modulated Gaussian demand with $\mu_{\mathrm{ON}}=1$\,kW. 
Shading indicates the four regions separated by the transition points $b_{\text{tx},1}$,
$b_{\text{tx},2}$ and $b_{\text{tx},3}$. 
\emph{(a)} high variance / long dwell ( $\sigma_1=158$\,W, $L=1000$ slots) (b) low variance / short dwell ($\sigma_1=32$\,W, $L=10$ slots). 
Curves: 
exact calculation using state$\times$battery Markov chain (black) and effective power $P_{\text{eff}}$ (red), 
both solid. The exact calculation is evaluated at logarithmically spaced battery sizes. 
Each region's approximation is drawn in a darker shade of that region's shading: 
Region 1a ideal bound (green, dashed), 
1b single state effective power (orange, dashed), 
2a linear mean reduction (blue, dashed), 
2b $\sigma_\text{eff}$ 2nd-order approximation (purple, dashed).}
\label{fig:MMPP}
\end{figure}

To gain architectural insight, consider the symmetric case and let  $L\triangleq L_1=L_2$.  In this case the system spends and equal amount of time in each state and the process mean and variance are $\mu_P = 0.5 \mu_1$ and $\sigma_p^2 = 0.5 \sigma_1^2 + 0.25 \mu_1^2$, respectively.  Two examples are shown in Fig. \ref{fig:MMPP}, one with a high ON state variance and long dwell time (1000 slots) and one with a low ON state variance and short dwell time (10 slots).  The exact power requirements are shown in black for battery sizes with margin reductions up to about $65-80\%$, above which numerical computation becomes burdensome. 

Both examples exhibit the SBR and LBR regions. In the SBR, power requirements fall linearly with $b$ until the SBR/LBR transition point, $b_{\text{tx}}$.  In Fig. \ref{fig:MMPP}, the SBR (Region 1a) is shown shaded in light green with the SBR approximation Eqn. (\ref{eqn:cap-sbr-mmp}) shown as dark green dashed lines. The approximation is quite accurate in the SBR, indicating the battery is almost always full and the system is operating at peak energy-to-power efficiency. The SBR approximation is plotted past the SBR/LBR transition point showing it is no longer valid in the LBR. In the LBR, the required power is approximately the effective power.  In Fig. \ref{fig:MMPP}, the LBR is the remaining portion of the graph, divided into 3 shaded subregions (Regions 1b, 2a, and 2b). The solid red line is the effective power, calculated from Eqn. (\ref{eq:MMPP_log_mmt}). As can be seen, the effective power is a very accurate measure of required power except during the SBR/LBR transition.

The four battery regions are best understood by dividing them into two qualitatively different phases: 
\begin{itemize}
    \item Phase 1 (Regions 1a, 1b): Here, the battery is large enough to reduce ON state power fluctuations, but not large enough to reduce the required power below the ON mean, $\mu_1$. In this region, $p_{r}(b)$ can be estimated by modeling the systems as always in the ON state, with an appropriately adjusted overdraw target. The region exhibits two subregions: the SBR region (R1a) and the first LBR subregion (R1b). In R1b, increased energy storage reduces the power requirement asymptotically towards $\mu_1$ until the 2nd transition point where the system enters Phase 2.  
    \item Phase 2 (Regions 2a, 2b): Here, the battery has effectively suppressed ON state variations and additional storage can be used to reduce the mean difference term of Eq. (\ref{eqn:margin-NB-MMP-gaussian}) by averaging power demands over multiple ON/OFF cycles. Phase 2 requires far more storage than Phase 1 but eventually drives the required power to the long term average $u_P = 0.5 \mu_{1}$.
\end{itemize}

Each region is described below in more detail. For each, we provide approximations which provide more architectural insight, at the slight cost of accuracy, than numerical calculations. For clarity, define $p_{r,1a}(b), p_{r,1b}(b), p_{r,2a}(b)$ and $p_{r,2b}(b)$ as the approximations to $p_r(b)$ in each region.

\subsubsection{Region 1a (SBR)}

In the SBR, $p_r(b)$ is is well approximated by Eq. (\ref{eqn:cap-sbr-mmp}), and thus
\begin{align}
p_{r,1a}(b) & =  \mu_1 + \alpha_1 \sigma_1 - b
\end{align}
where $\alpha_1 = 2.84$ for $\varepsilon = 10^{-3}$, indicating slightly less power per standard deviation is needed compared to an equivalent system that is always ON ($\alpha = 3.05$). Even though the system spends only half its time in the ON state, the required margin without a battery is only reduced by $1 - \alpha_1 / \alpha \approx  7\%$.

The SBR region continues until the 1st transition, the SBR/LBR transition, $b_{\text{tx},1} = b_{\text{tx}} \approx 0.5 \alpha_1 \sigma_1$. At this point, the power has been reduced to the first midpoint, $p_{r,1a}(b_{\text{tx},1}) = u_1 + 0.5 \alpha_1 \sigma_1$ and multi-slot overdraws become significant. 

Fig. \ref{fig:MMPP} shows Region 1a in green. $p_{r,1a}(b)$ is shown as a green dashed line indicating the accuracy of the approximation and that in R1a, the margin requirements are primarily determined by the ON state statistics. 

\subsubsection{Region 1b (ON state LBR)}

Here, ON state variations still dominate and the required power is well approximated by the effective power of the ON state with an adjusted overdraw probability. Thus, 
\begin{equation}
p_{r,1b}(b) = \mu_1 + \frac{\delta_1 \sigma_1^2}{2b}
\end{equation}
which asymptotically approaches $\mu_1$ until the benefits of energy storage saturate.  Saturation  occurs when the battery size is large enough that overdraw times start to exceed ON state dwell times. This battery size, $b_{\text{tx},2}$ can be estimated as point where the typical overdraw time for an always ON process, $T_{\text{typ}} \approx L$, and thus since $T_{\text{typ}} \approx \frac{2b^2}{\delta_1 \sigma_1^2}$, 
\begin{align}
b_{\text{tx},2} = \sqrt{\frac{1}{2} L\delta_1 \sigma_1^2} 
\end{align}
Thus, Region 1b is defined as the battery sizes $b \in \left[0.5 \sigma_1 \alpha_1, \sigma_1 \sqrt{0.5 \delta_1 L}\right]$. The required power at the end of Region 1b is 
\begin{equation}
p_{r,1b}\left(b_{\text{tx},2}\right) = \mu_1 + \sigma_1 \sqrt{\frac{\delta_1}{2L}} \label{eqn:R1b-2nd-order}
\end{equation}
The longer the ON dwell time and the smaller the ON state variations, the larger the region and the closer $p_{r,1b}(b)$ gets to $\mu_1$.  For long dwell times, the region suppresses nearly all the ON-state variations and the required power is approximately the ON state mean.

In Fig. \ref{fig:MMPP}, 
Region 1b is shaded light orange with $p_{r,1b}$ is shown as a dashed dark orange line. 
The approximation matches the effective power approximation over the entire region and the exact calculation quite well over a large portion of it. However, both approximations are slightly inaccurate in the SBR/LBR transition, as seen previously for single-state processes. 

Note that the ON dwell time need not be that large for Region 1b to be significant.  For instance, the typical overdraw time (at $\varepsilon = 10^{-3}$) for a single state Gaussian i.i.d. process is only about 6 slots at 75\% margin reduction. Thus, for most parameters of interest, Region 1b exists and reduces margin requirements significantly. On the other hand, for extremely small ON dwell times, i.e. on the order of a single slot, Region 1b can be small to essentially non-existent. For example, if $L = 2$, then long sojourns in the ON state which could slowly drain the battery do not exist because frequent OFF transitions recharge the battery. Recall that the battery being almost full is a hallmark of the SBR.

\subsubsection{Region 2a - Mean Reduction, part a}

Region 2a starts the beginning of Phase 2, where having effectively suppressed ON state fluctuations, additional energy storage can reduce the required power below the ON mean. This mean reduction phase occurs over two sub regions.  The first, Region 2a is characterized by the fact that with very high probability each ON period starts with a fully charged battery. It ends when that is no longer true.

In Region 2a, to a rough approximation, $b$ units of \textit{additional energy} can contribute $b/L$ units of power for mean reduction below the ON state. However, dwell times are random and this deterministic approximation is inaccurate. Using a simple model where the ON state is modeled with constant demand $u_1$, the system always starts the ON state with a full battery, $b$, and it stays in the ON state for a geometrically distributed amount of time, it can be shown that
\begin{align}
p_s \geq u_1 + \frac{b}{\delta_1} \log{\left(1 - \frac{1}{L} \right) }
\end{align}
is a necessary condition to maintain a target overload [Appendix \ref{app:region2a}]. Adding this requirement to the Region 1b requirement, we have
\begin{equation}
p_{r,2a}(b) \approx u_1 + \frac{\delta_1 \sigma_1^2}{2b} + \frac{(b-b_{\text{tx},2})}{\delta_1} \log{\left(1 - \frac{1}{L} \right) } \label{eqn:on-off-2a}
\end{equation}
Thus, in R2a, the required power again falls linearly, but here with a greatly reduced slope of $s = \frac{1}{\delta_1} \log{\left(1 - \frac{1}{L} \right) } < 0$.  Region 2a ends at $b_{\text{tx},3}$ when there is insufficient time to reliably fully charge the battery during an OFF period. Since there are no power demands in the OFF state, there is approximately, on average, $p_r(b) L$ power charge available to charge the battery. Thus, the transition point can be estimated by solving $b_{\text{tx},3} = p(b_{\text{tx},3})L$.  Since $p_r(b) \approx \mu_1 - sb$, it follows that 
\begin{align}
b_{\text{tx},3} & \approx \frac{L u_1}{1 - L s}
        \approx \frac{\delta_1}{\delta_1+1} u_1 L \\
        p(b_{\text{tx},3}) & \approx \mu_1 \left(1 - \frac{\delta_1}{\delta_1 +1} \right) 
\label{eqn:on-off-recharge}
\end{align}
For $\varepsilon = 10^{-3}$, this corresponds to the prediction that at the end of the region, the mean has been reduced about $14\%$ below $\mu_1$; for $\varepsilon = 10^{-2}$, used in the next section, the predicted reduction is about $20\%$.

Fig. \ref{fig:MMPP} shows Region 2a for the two examples shaded as light blue. The approximation $p_{r,2a}$, shown as a dark blue dashed line, lines up very well with the exact Markov calculations and the numerical computation of effective power. For the two examples, exact calculations shows the system ends the region at $13.2\%$ ($L = 1000$) and $14.2\%$ ($L = 10$) below the mean.  The reduction in R2a decreases with the overflow target, e.g. at $\varepsilon = 10^{-6}$, R2a ends at about $6.7\%$ below the mean with the above approximation accurate to within $1\%$ for the two examples and for $\varepsilon = 10^{-2}$ and $10^{-6}$. 

\subsubsection{Region 2b - Mean Reduction, part b}

After $b_{\text{tx},3}$, additional storage continues to reduce the mean to the long term average but with diminishing returns. 

For very large battery sizes, well into Region 2b, the remaining margin can be approximated using the second order expansion on $P_{\text{eff}} \approx \mu_P + \frac{\delta \sigma_{\text{eff}}^2}{2b}$, where in general, 
\begin{align}
    \label{eq:on-off-cov-2}
\sigma_\text{eff}^2 =\frac{\mathcal{P}_{21}}{\mathcal{P}_{12}+\mathcal{P}_{21}}\sigma_1^2+u_1^2 \frac{\mathcal{P}_{12}\,\mathcal{P}_{21}}{(\mathcal{P}_{12}+\mathcal{P}_{21})^2} \frac{1+\gamma}{1-\gamma}
\end{align}
and where $\gamma=1-\mathcal{P}_{12}-\mathcal{P}_{21}$. As a sanity check, as $\mathcal{P}_{21}\to 1$ and $\mathcal{P}_{12}\to 0$, then $\sigma_\text{eff}^2 
\to \sigma_1^2$ as expected. For the symmetric case, Eq.~(\ref{eq:on-off-cov-2}) simplifies to $\sigma_\text{eff}^2 = \sigma_P^2 + \frac{1}{4} \mu_1^2 (L-2)$ and we have the R2b regional approximation
\begin{align}
    \label{eq:on-off-2b}
    p_{r,2b}(b) & \approx \mu_P + \frac{\delta}{2b} \left(\sigma_P^2 + \frac{1}{4} \mu_1^2 (L-2) \right)
\end{align}
In the special case where $L=2$, the process $P(t)$ is i.i.d. and $\sigma_{\text{eff}}^2 = \sigma_P^2$.  As the dwell time increases,  temporal correlations increase, and thus the effective power increases with $\mu_1^2 L$.  As expected the longer the dwell time, or the higher the ON mean, the higher the effective power and larger and larger batteries are needed to drive the power requirement to the long term mean. 

Region 2b is the last region in Fig. \ref{fig:MMPP}. Unlike the other per-region approximation, the $\sigma_{\text{eff}}$ approximation is only accurate for fractional margins of $\approx 20-30\%$ or less, well into the region . Before that, the full effective power calculation should be used. 

\subsection{Discussion}

This section demonstrated that the SBR (ideal)/LBR (effective power) formulation applies to a wider set of demand statistics than the previous two sections demonstrated, and that the LBR can exhibit multiple regimes with changing energy-to-power conversion efficiencies.

Since margins are invariant to shifts in the mean, the ON/OFF analysis would apply equally well to HIGH/LOW demands where the states are well separated. We will use this formulation to model real demand data in the next section. 

%% file: data_analysis.tex
\section{Data Integration and Analysis}
\label{sec:evaluation}

We evaluate the proposed framework using three production data-center data sets spanning substantially different workload types, operator scales, and measurement methods: batch-scheduled HPC workloads from the Ohio Supercomputer Center (OSC), long-running GPU training jobs from the Microsoft Philly cluster (Philly), and user-facing cloud services from (Alibaba). Despite these differences, we show that these real data-center power demands exhibit a common multi-timescale structure: relatively fast fluctuations occur within short-time-scale demand bursts, while transitions across these bursts introduce variations over substantially longer time scales. 

Of particular importance to power provisioning are two characteristics of the HIGH-demand bursts: the fast demand fluctuations that occur within each burst that account for most overdraw events and the frequency with which these HIGH-demand bursts occur. The former determines how effectively storage can absorb short-time-scale demand variations within a HIGH burst, while the latter determines the longer-time-scale opportunity for the battery to recover and average demand across bursts. These two distinct dynamics govern the required provisioned power at different storage scales, naturally separating the required-power-versus-storage curve into two phases. This two-phase behavior is naturally modeled by the simple two-state HIGH/LOW Markov model developed below. While not intended to reproduce the complete statistical structure of real data-center demand, the model captures the essential intra-burst fluctuations and inter-burst dynamics that determine how storage translates into reductions in required provisioned power.

Our objective in this section is not to precisely characterize the boundaries between different regions of operation, but rather to demonstrate their presence in measured data and show that the simple models developed in Section~\ref{sec:correlated} capture the fundamental operational dynamics that govern the storage-power tradeoff in real data-center loads.


\subsection{Data Sets}
\label{sec:data sets}


\noindent \textbf{OSC:} The Ohio Supercomputer Center (OSC)~\cite{osc2024} operates several
production HPC systems at The Ohio State University. The OSC trace logs
rack-level power drawn every $30$ seconds at $74$ rack-level Power Distribution Units (PDUs) in the center's production facility. 
The compute load comes from $48$ computing racks across three clusters,
\emph{Ascend} (AMD EPYC CPUs with NVIDIA A100 GPUs),
\emph{Cardinal} (Intel Xeon CPU~Max nodes with NVIDIA H100 GPUs), and \emph{Pitzer}
(Intel Xeon Skylake/Cascade~Lake nodes, a subset with NVIDIA V100 GPUs), 
and the remaining $26$ racks cover supporting storage and infrastructure.

Power is measured directly at rack PDUs every $30$\,sec by the center's facility power monitoring system and historical data was downloaded through the center's Prometheus metrics API with access provided by the data center. Unlike the Philly and Alibaba traces, this data set is manually collected and not publicly available.

The per-rack power traces were aligned to a common $30$-second grid and summed to form the aggregate group-power process $P(t)$. The resulting trace spans six days (March 13--18, 2026) and contains $17{,}279$ samples. Since these are physical readings, they capture the full server-plus-infrastructure draw of the rack. 

\noindent \textbf{Philly:} The Microsoft Philly trace~\cite{jeon2019atc, philly2017trace} logs per-minute GPU
utilization for a large multi-tenant production cluster used for deep neural network
(DNN) training over a $22.6$-day period (October 3-25, 2017). The cluster mixes two server types: 2-GPU servers (NVIDIA K80/M40-class, $12$\,GB per GPU) and $8$-GPU servers (NVIDIA P40-class, $24$\,GB per GPU).
Of the $552$ servers in the trace, we use the $420$ with complete coverage over the period. The per-GPU utilization traces were aligned to a common $60$-second grid. To estimate a server's power demand $X(t)$ from it's GPU utilization $u(t)$, we first estimate the power draw from each of its GPU's using a simple linear model~\cite{fan2007},  $G(t) =G_{\text{idle}} + (G_{\text{TDP}} - G_{\text{idle}}) u(t) $ where $G_{\text{idle}}$ is the GPU's idle power draw and $G_{\text{TDP}}$ is the GPU's total design power. These are then summed together and added to a per server baseline draw to estimate $X(t)$. For the $2$-GPU servers, the server baseline was set at $150$\,W and the idle/TDP draws were $50/150$W. For the 8-GPU servers, the server baseline was set at $300$\,W and the idle/TDP draws were $30/250$\,W.  The estimated $420$ server power draws were then aggregated to form a group-power process $P(t)$ with $32,521$ samples. The trace is publicly available~\cite{philly2017trace}.

\noindent \textbf{Alibaba:} The Alibaba production cluster trace 
(v2018)~\cite{alibaba2018trace}
logs per-server CPU utilization over $8$ days for a cloud cluster of $3,989$ 
heterogeneous servers that co-locate latency-sensitive online services with batch
jobs. Due to incomplete data in the first two days, we analyzed the data over days $3$-$8$.  
The per-server utilization traces were aligned to a common 60-second
grid. To estimate a server's power demand from it's CPU utilization $u(t)$, we first estimate the power draw from its CPUs using the same linear model, where $C(t) =C_{\text{idle}} + (C_{\text{TDP}} - C_{\text{idle}}) u(t) $ where $C_{\text{idle}}$ is the CPU's idle power draw and $C_{\text{TDP}}$ is the CPU's total design power.  $C_{\text{idle}}$ and $C_{\text{TDP}}$ were assumed to be $3$\,W and $10$\,W, respectively per core of the CPU. So, e.g., an $8$-core CPU would have idle/draw parameters $24$\,W$/80$\,W. This is then summed together and added to a per server baseline draw of $100$\,W to estimate $X(t)$. The $3,989$ server demands were then aggregated to form the group-power process $P(t)$ with $8,640$ samples. The trace is publicly available~\cite{alibaba2018trace}.

Table~\ref{tab:data sets} summarizes the aggregate group-power process $P(t)$ for each facility. The three data sets span an order of magnitude in scale in mean power demand ($288$ to $2{,}841$\,kW) from $74$ to $3{,}989$ sources. All three single slot power distributions are multi-modal with the number of modes varying from $2$ to $5$, and skews varying from $-1.06$ (left) to ($+0.62$) (right). In terms of variation around the mean, the coefficient of variations (CV), $\sigma_P/u_P$, differ by about a factor of $2$ from $0.05-0.11$. The $0.01$ quantile powers differ, resulting in a range of $1.6-2.6$ burst sizes, i.e. standard deviations of required margin, to meet an overdraw target of $0.01$ without any energy storage. 

The three traces are also strongly temporally correlated with very high single slot correlations and exhibit indications of multiple time scales. Long term correlations can be measured by the \textit{ correlation time}, the lag at which the autocorrelation of $P(t)$ first falls to $0.5$. The long term time scales vary by about a factor of $4.4$ with Alibaba exhibiting the least long term temporal correlation of  $127$\,min, consistent with short-lived cloud jobs, Philly exhibiting the longest $561$\,min consistent with long-run training, and OSC sitting firmly in the middle at $328$ min consistent with its mixed use nature. Since power margin requirements without energy storage are solely determined by the tails of the distributions, Table~\ref{tab:data sets} lists the average \textit{burst persistence} at $0.01$ quantile. Here, the temporal order is reversed with Philly exhibiting the least persistence ($2$ min $= 2$ slots), Alibaba the most persistence ($7$\,min $= 7$\,slots), with OSC in the middle again at ($3.5$\,min $= 7$\,slots).  Although OSC has the same slot persistence as Alibaba, it also exhibits the most skewed distribution of burst lengths with a maximum burst of $58$\,slots compared to Alibaba's maximum burst length of $15$ and Philly's of $10$.

\begin{table}[t]
\centering
\caption{Summary of the three real-world data sets.}
\label{tab:data sets}
\renewcommand{\arraystretch}{1.15}
\footnotesize
\begin{tabular}{@{}lccc@{}}

\toprule
& \textbf{OSC} & \textbf{Philly} & \textbf{Alibaba} \\
\midrule

Operator     & Ohio State & Microsoft & Alibaba \\
Type         & HPC & GPU training & Cloud \\
Measurement  & Direct (PDU) & Est. (GPU)& Est. (CPU)\\
Resolution   & 30 sec& 60 sec& 60 sec\\
Duration     & 6 days & 22.6 days & 6 days \\
\# of Samples & 17{,}279 & 32{,}521 & 8{,}640 \\
Sources      & 74 racks & 420 servers & 3,989 machines \\
Mean ($u_P$) & 883.8 kW& 287.7 kW& 2,841 kW\\
 Std. Dev. (kW) ($\sigma_P$)& 50.4 kW& 15.1 kW&298.9 kW\\
Coefficient of variation & 0.057 & 0.053 & 0.105 \\
Skewness     & $-0.06$ & $-1.06$ & $+0.62$ \\
 Number of modes& 4-5& 4&2\\
 .01 Quantile& 981.9 kW& 312.6 kW&3609.4 kW\\
1-slot correlation & 0.997 & 0.978 & 0.949 \\
 Correlation Time& 328 min& 561 min&127 min\\
 .01 Quantile persistence& 3.5 min& 2 min&7 min\\
 \bottomrule
\end{tabular}
\end{table}

\subsection{2-State High/Low Model}
\label{sec:high-low}

The demands in each of the three data sets are complex showing multi-state multi-time scale cyclo-stationary behavior with high temporal correlations. However, based on the insights developed for symmetric ON/OFF processes (Section \ref{sec:on-off}), we expect the required margin to be primarily dependent on the worst case demand statistics and a coarse summary of the statistics when not in the worst case state. 

To test this, we modeled each data set using a simple 2-state HIGH/LOW process with Gaussian i.i.d. demands in each state distributed as $\mathcal{N}(\mu_{H},\sigma_H)$ and $\mathcal{N}(\mu_{L},\sigma_L)$. The process dwells in the HIGH and LOW states for an average of $L_H$ and $L_L$, respectively. Since margin requirements are mainly a function of the HIGH state and its separation from the LOW state, and less so on state dwell times, we make the simplifying assumptions that $L_H = L_L$ and $\sigma_H = \sigma_L$, resulting in a 4-parameter demand model. The process we used to determine the model parameters from the data sets is laid out in Appendix \ref{app:model_params}.  

The resulting demand parameters are given in Table \ref{tab:2-state-model}. Here, consistent with the demand statistics, Philly and OSC exhibit similar long correlation times (in slots), with Alibaba's dwell times much smaller. Clearly, the $4$ parameters are insufficient to statistically describe complex power demands; however, that is not our goal. Rather, the goal is to predict the achievable power savings from energy storage.  Given the accuracy of this model, as outlined in the next section, these 4 parameters better predict the behavior of the system than the statistical properties laid out in Table \ref{tab:data sets}.

\begin{table}[t]
\centering
\caption{Two-State Gaussian Model}
\label{tab:2-state-model}
\renewcommand{\arraystretch}{1.15}
\footnotesize
\begin{tabular}{@{}lrrr@{}}
\toprule
& \textbf{OSC} & \textbf{Philly} & \textbf{Alibaba} \\
\midrule
High Mean      & 953.8 kW & 308.6 kW & 3400 kW   \\
High Std. Dev. & 14.19 kW & 2.01 kW & 105.64 kW \\
Dwell Time     & 2.0 hrs & 4.3 hrs & 0.4 hrs   \\
               & (236.8 slots) & (259.3 slots) & (23.9 slots)      \\
Low Mean       & 813.8 kW & 266.8 kW & 2282 kW   \\
Low Std. Dev.  & 14.19 kW & 2.01 kW & 105.64 kW \\
\bottomrule
\end{tabular}
\end{table}

\begin{table}[t]
\centering
\caption{Two-state Model: battery size $b_e$ (kWh) at the end of each region and simulation.}
\label{tab:2-state-model-regions}
\renewcommand{\arraystretch}{1.15}
\setlength{\tabcolsep}{5pt}
\footnotesize
\begin{tabular}{|@{}c|c|c|c|c@{}|} \hline
 & \multicolumn{2}{|c|}{\textbf{Phase 1}}& \multicolumn{2}{|c|}{\textbf{Phase 2}}\\\hline
 & \textbf{SBR 1a end}& \textbf{1b end}& \textbf{2a end}& \textbf{sim. end}\\ \hline
 & \textbf{$b_{\text{tx},1}$} & \textbf{$b_{\text{tx},2}$}& \textbf{$b_{\text{tx},3}$} & \textbf{} \\ \hline 
               
OSC     & 0.050 & 46.7 & 179 & 179\\\hline
Philly  & 0.033 & 4.83 & 153 & 1667\\\hline
Alibaba & 2.33 & 70 & 267 & 1817\\\hline

\end{tabular}
\end{table}

\begin{figure*}[!tp]
\centering
\includegraphics[width=0.95\textwidth]{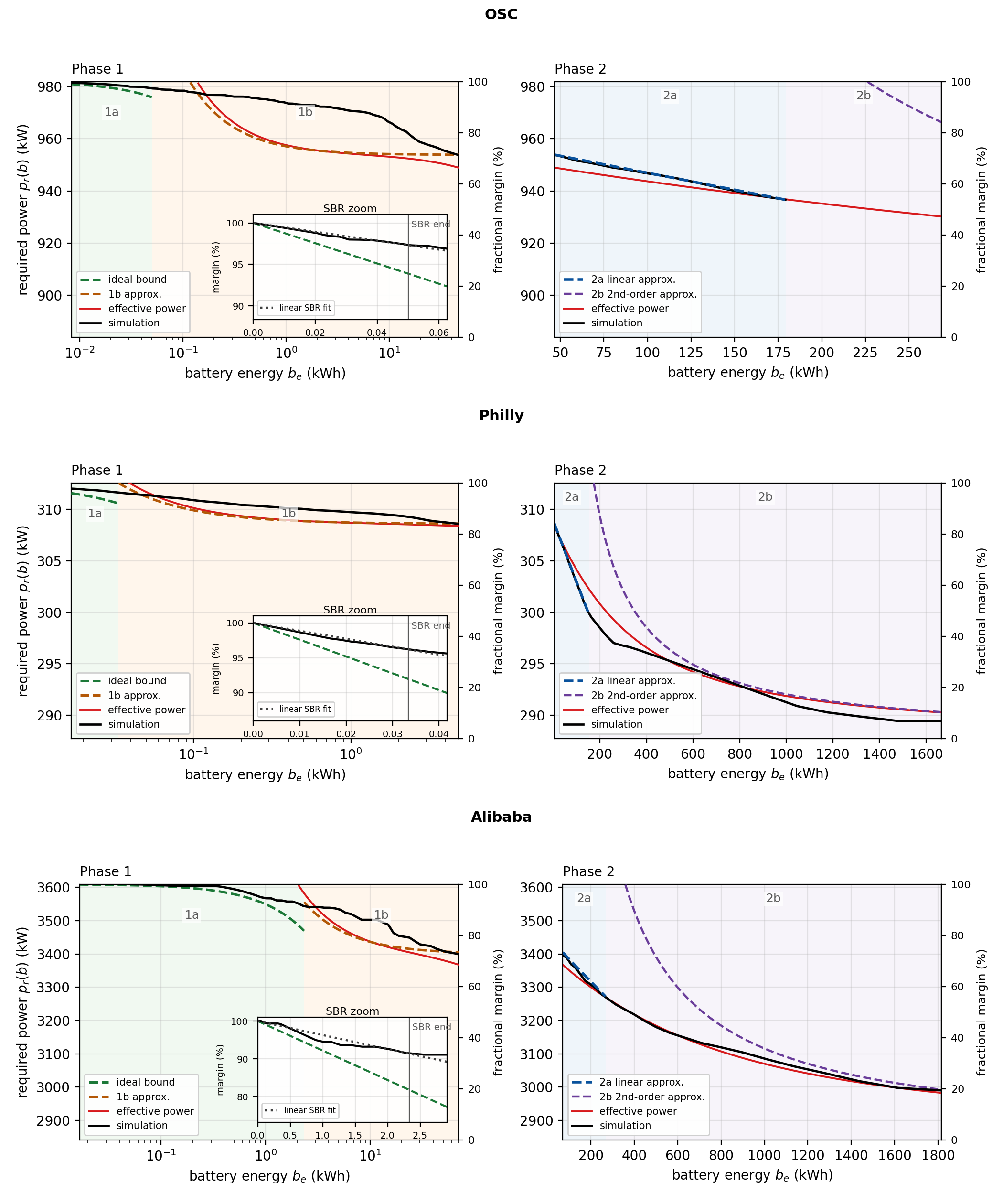}
\caption{Required power $p_r(b)$ (left axes) and fractional margin (right axes) vs.\ battery energy $b_e$ for an overflow target of $\varepsilon = 0.01$.  Top to bottom: OSC, Philly, and Alibaba data sets with each divided into two panels. 
Left panel: Phase 1 divided into its regions R1a/R1b.
Right panel: Phase 2 divided into its regions R2a/R2b.  
Curves: empirical simulations (solid black) and LBR $P_\text{eff}$ approximation (red).  Shading indicates the four regions over their empirical transition points. The regional approximations are shown in each: R1a SBR approx. (green dashed), R1b approx. (dashed orange), R2a approx. (dashed blue), R2b approx. (dashed purple).
The left panel also contains a zoomed inset of R1a showing a fitted linear line (dotted black). For OSC, the simulation ends at $b_{\text{tx},3}$, so only the model curves are shown in R2b.}
\label{fig:realdata-capacity}
\end{figure*}

\subsection{Data Driven Simulations}
\label{sec:data-drive-simulations}

To assess whether the fundamental operational regimes identified by our models are reflected in real data-center workloads, we simulated the system in Fig.~\ref{fig:single-layer} with different battery sizes using the OSC, Philly, and Alibaba data sets. Recall that, in our conceptual development, we have used $p_r(b)$ as the required power, where $b$ is the battery capacity, measured in units of power charge of the battery, which is the energy capacity, normalized per time slot. In this section, we will present our results in core physical quantities; thus we will report $p_r(b_e)$, where $b_e$ is the raw battery capacity, measured in Watt-hours (Wh). 

We simulate the system using the power charge battery state equations (\ref{eq:battery-update-energy}) to determine the required power $p_r(b_e)$ to meet the overdraw target as a function of the battery power charge capacity, $b$. Given the limited number of samples, the overflow target was set at $0.01$. For each battery size, $b$, $p_r(b)$ was determined using a bisection method to a resolution of $1$\,W. Our analysis is aligned with a potentially large group of racks, supported by a centralized energy storage unit such as a BESS. Note that, the load can potentially be distributed and supported by a distributed group of racks, each with their individual energy storage unit. 

In each iteration, the simulation started with an empty battery state. This provides a conservative estimate of $p_r(b)$ and has the added benefit that all simulations start identically. To eliminate start-up effects, each set was simulated a second time starting with a full battery in each iteration.  The start-empty and start-full required powers only differ for very large battery sizes. The simulation endpoints, defined as the battery size, where the two methodologies differed by more than $2.5$ percentage point of fractional margin, are shown in the last column of Table~\ref{tab:2-state-model-regions}.  For Alibaba and Philly, the start-up battery state was largely irrelevant with the simulations covering $\sim 80\%$ and $\sim 93\%$ of potential margin savings, respectively. However, for OSC, the empty/full battery simulations diverged significantly sooner and thus only the first $46\%$ of potential savings are simulated, limiting our ability to test the model on OSC for larger battery sizes where the battery is being used to average over many demand cycles. This limitation is due to an insufficient number of demand cycles in the OSC data set, which itself is due to an insufficient time period for data with long busy periods. 

Each system was simulated for $100$ different values of the power charge $b$, geometrically spaced between $b = 1$\,kW-slot and the simulation stopping points. The results were then converted to energy using the relationship $b_e = bT_s$, where we set the time slot size to the data sample length ($T_s = 30$ sec for OSC and $60$ sec for Philly and Alibaba). Note with a slight abuse of notation, we will use $p_r(b)$ and $p_r(b_e)$ to represent the required power as a function of battery power charge or battery energy, respectively. 

Fig.~\ref{fig:realdata-capacity} shows the results of the simulations for each data set as well as the analytical predictions of the HIGH/LOW model with the SBR ideal bound/approximation shown in dashed green and the LBR $P_\text{eff}$ approximation shown in solid red.  The empirical curve stops at the simulation stopping points. This coincides with the end points of the graphs for Philly and Alibaba. For OSC, we continue the graphic past the empirical point.

Power savings occur over the two phases identified in Section~\ref{sec:on-off} for ON/OFF Gaussian i.i.d. models. In Phase 1 (left panels), energy storage is used to suppress HIGH state variations, driving $p_r(b_e)$ towards the HIGH state mean, $u_H$, resulting in margin savings of about $30\%$ (OSC), $18\%$ (Philly), and $25\%$ (Alibaba).  In Phase 2 (right panels), energy storage has largely suppressed HIGH state bursts and any additional energy storage reduces the required power below the high state mean toward the long term average of $u_p$. 

Each phase is further divided into a linear sub region (R1a and R2a) where the battery's marginal efficiency is near constant, and a sublinear region (R2a and R2b) where the battery's marginal efficiency reduces with battery size. The sub regional boundaries are determined empirically as described in Appendix~\ref{app:model_params} and shown in Table~\ref{tab:2-state-model-regions} in kWh. Each region is shaded in Fig.~\ref{fig:realdata-capacity} and includes its regional approximation: Eq.~\ref{eqn:capacity-SBR}) for SBR (1a),  Eq. (~\ref{eqn:R1b-2nd-order}) for the first LBR region R1b, Eq.~(\ref{eqn:on-off-2a}) for the mean reduction region R2a, and Eq.~(\ref{eq:on-off-2b}) for the diminishing return region of Phase 2. 

In general, the model prediction errors are small, especially for Philly 
and Alibaba. In addition, regional approximations for 1b and 2a are very close to the $P_\text{eff}$ calculations with some small additional inaccuracy at the start of 1b and the end of 2a, confirming the regional physical insights developed in Section~\ref{sec:on-off}. As pointed out in Section~\ref{sec:on-off}, the R2b regional approximation is only expected to be accurate well into region $2b$.

Overall, the Philly data matches the closest with the 2-state abstraction, with a difference under $5\%$ throughout the curve, except near the R2a/b transition where the error is closer to $10\%$. The 2-state model fits Alibaba model accurately in Phase 2, but slightly less accurately overall with a maximum error of less than $10\%$ at the SBR/LBR (R1b) transition as well as in mid R1b. In terms of maximum error with respect to the 2-state model, OSC data is the least accurate overall, with maximum prediction errors of about $3\%$ in the SBR and about $17.5\%$ in mid R1b. 

At the boundaries of the regions, the model is more accurate across all data sets, as can be seen in Table \ref{tab:prediction-errors}. At the regional transition points, the error is typically $5\%$ or less except for two outlying cases of about $10\%$ error. 

From an operational perspective, the predictions differ from the empirical data in several significant ways. Although each data set exhibits the four predicted sub regions, the predicted battery transition points presented in Section~\ref{sec:on-off} were not accurate. As a result, we reported the empirically observed transition values. For instance, the SBR region differed from expectations in both 1) a marginal battery efficiency of roughly half of what is predicted and 2) the system transitioning out of the SBR sooner than expected. The reduced efficiency can be seen in left panel (see the inset zoom). The system exhibits the expected linear behavior with a dashed linear fit line\footnote{the fit is simply the straight line 
connecting $p(0)$ and $p_r(b_{\text{tx},1})$}, however with slopes roughly half of ideal, 
specifically $53$ kW/kWh for OSC and $28$ kW/kWh for Philly and Alibaba, versus ideal slopes of $120$kW/kWh and $60$ kW/kWh, respectively.

We attribute these differences to temporal correlations not considered in Section~\ref{sec:SBR}. To test this theory, we re-simulated each data set $30$ additional times with the power demands permuted in time. The average ``i.i.d.'' efficiencies 
were within $2.2\%$ of the ideal bound,
indicating that the reduced SBR slopes are primarily due to temporal correlations. The transition of the system out of the SBR is also not predicted well by the Gaussian i.i.d. point of $b_{\text{tx},1}  \approx 0.5 \alpha_1 \sigma_1$. The differences are likely attributable to both intra-state correlations and HIGH state power demands differing from Gaussian, as will be discussed in Section~\ref{sec:sbr-discussion}.  The combination of these two effects results in mismatches with what is predicted with the 2-state model: $2.7\%$ actual v. $7.2\%$ predicted for OSC,  $3.8\%$ actual v. $4.0\%$ predicted for Philly, and $8.6\%$ actual v. $6.8\%$ predicted for Alibaba. The biggest difference being for OSC where the SBR produces only about $30\%$ of its potential savings.  

In summary, the empirical data confirms that margin reduction occurs over two phases, each divided into two sub-regions. The prediction accuracy of the 2-state model was good, but the predicted transition points had to be determined empirically. The remaining inaccuracies are likely mainly attributable to unmodeled HIGH state temporal correlations.

\begin{table}[t]
\centering
\caption{Predicted minus actual fractional margin, in percentage points, at the end of each region.}
\label{tab:prediction-errors}
\renewcommand{\arraystretch}{1.15}
\setlength{\tabcolsep}{5pt}
\footnotesize
\begin{tabular}{|@{}c|c|c|c|c@{}|} \hline
 & \multicolumn{2}{|c|}{\textbf{Phase 1}}& \multicolumn{2}{|c|}{\textbf{Phase 2}}\\\hline
 & \textbf{SBR 1a end}& \textbf{1b end}& \textbf{2a end}& \textbf{sim. end}\\ \hline
 & \textbf{$b_{\text{tx},1}$} & \textbf{$b_{\text{tx},2}$}& \textbf{$b_{\text{tx},3}$} & \textbf{} \\ \hline
OSC     & $-3.4$ & $-5.0$ & 0.2 & -- \\\hline
Philly  & $-4.2$ & $-0.8$ & 9.9 & 3.4 \\\hline
Alibaba & $-9.6$ & $-4.2$ & $-0.1$ & $-1.0$ \\\hline
\end{tabular}
\end{table}

%% file: conclusion.tex
\section{Conclusion and Future Work}
\label{sec:conclusion}

This paper developed a probabilistic theory for power provisioning in data centers equipped with distributed energy storage. By viewing the battery as a stochastic resource that absorbs demand fluctuations across time, we established a unified framework that directly relates provisioned power, battery capacity, and target overdraw probability. 

The analysis revealed that the provisioning problem naturally decomposes into two fundamentally different operating regimes. In the Small Battery Region, the battery can compensate only short-lived demand excursions, so the required provisioned power is dominated by the upper tail of the instantaneous demand distribution. Consequently, the power margin above the average demand decreases linearly with increasing battery capacity, with the size of the region governed primarily by the variance, tail behavior, and temporal correlations of the workload. In contrast, in the Large Battery Region, sufficiently large storage smooths out longer-term fluctuations, allowing the provisioned power to approach the long-term average demand. The remaining margin is no longer determined by instantaneous peaks, but by the effective power of the workload, which captures the cumulative stochastic behavior of the demand over extended time horizons. As a result, additional storage yields progressively diminishing reductions in the required provisioned power, and the asymptotic provisioning limit is dictated by the long-term statistical properties of the demand rather than its instantaneous variability.

Together, these two regimes provide a complete picture of how distributed energy storage converts temporal variability into reductions in required supply capacity. Beyond the analytical development, the framework provides practical guidance for battery dimensioning, identifies the onset of diminishing returns as storage increases, and quantifies how storage and statistical multiplexing jointly determine provisioning efficiency. 

In addition, the paper develops a comprehensive statistical model of data-center workloads incorporating temporal correlations across multiple time scales and establishes that a small number of dominant parameters determine energy/power provisioning trade-offs.

Finally, the theoretical framework is validated using power measurements collected from three production data centers. Despite the complexity and multi-modal nature of these workloads, the measurements demonstrate that the proposed stochastic models accurately capture the dominant operational behavior and provide reliable predictions of the provisioning tradeoffs, illustrating that relatively simple probabilistic abstractions are sufficient to obtain meaningful engineering insights for practical system design.

The framework developed here also provides the foundation for a broader theory of hierarchical energy management. Future work will extend the single-layer abstraction to multi-layer distributed energy architectures in which fast storage technologies, such as supercapacitors, operate at lower levels of the hierarchy while higher-capacity battery systems serve larger collections of racks and clusters. Such architectures naturally introduce multiple temporal and spatial scales, requiring a joint optimization of power allocation across layers based on the effective power of aggregated workloads. Developing this multi-scale theory, together with practical control algorithms for heterogeneous energy-storage systems, represents a promising direction toward systematic, statistically provisioned power infrastructures for next-generation AI data centers.

%% file: MC.tex
\section{Markov Chain Analysis}
\label{app:markov}

%
%

When the aggregate demand $P(t)$ is i.i.d., the battery recursion~\eqref{eq:battery-update} makes the charge level $b(t)$ a time-homogeneous Markov chain. The overdraw rate is then computed exactly as a stationary average rather than estimated by simulation. The construction below applies to a general demand distribution and is specialized afterward to the integer (Poisson) and continuous (Gaussian) cases.

\subsection{Discretizing the charge level}
\label{app:markov-discretization}

The charge level $b(t)$ is continuous on $[0,b]$, so a finite chain is obtained by representing it with a grid of $k+1$ states $\{0,1,\ldots,k\}$ at spacing
\begin{equation}
    h = \frac{b}{k}.
    \label{eq:mc-resolution}
\end{equation}
State $i$ corresponds to the charge $b_i = i\,h$, and more precisely to the cell of charges within $h/2$ of $b_i$. Because the grid is defined as a fraction of $b$, the resolution is relative rather than absolute. A choice of $k = 100$ resolves the battery to one percent of its capacity, regardless of the physical units in which $b$, $p_s$, and $P(t)$ are expressed. The unit of power sets neither the number of states nor the accuracy, and only $k$ does. The discretization error decreases as $k$ grows, and $k$ is taken large enough that the computed overdraw rate has converged.

When the demand is integer valued, the choice $h = 1$ gives $k = b$ and makes state $i$ exactly the charge $i$. The cells are then the integers and no approximation is introduced, which is the natural setting for the Poisson model.

\subsection{Transition probabilities}
\label{app:markov-transitions}

From~\eqref{eq:battery-update}, the next charge given $b(t) = b_i$ is the clamped value $\hat b = \clamp{p_s + b_i - P(t)}{0}{b}$. State $i$ moves to state $j$ when $\hat b$ rounds to the cell of $b_j$, which happens when the demand falls in an interval of width $h$ centered at $p_s + b_i - b_j = p_s + (i-j)\,h$. The two boundary states absorb the mass that clamps to an empty or full battery. The transition matrix $\mathbf{P} = [P_{ij}]$ collects the resulting state-to-state probabilities,
\begin{equation}
    P_{ij} =
    \begin{cases}
        \Prob \big[\,P(t) > p_s + b_i - \tfrac{h}{2}\,\big], & j = 0,\\[4pt]
        \Prob \big[\,\big|\,P(t) - (p_s + (i-j)h)\,\big| \leq \tfrac{h}{2}\,\big], & 0 < j < k,\\[4pt]
        \Prob \big[\,P(t) \leq p_s + b_i - b + \tfrac{h}{2}\,\big], & j = k.
    \end{cases}
    \label{eq:mc-transition-general}
\end{equation}
The interior entries depend only on the difference $i-j$, so $\mathbf{P}$ has a banded Toeplitz structure and is inexpensive to build.

For integer valued demand with $h = 1$, the interior mass is the demand pmf $p(p_s+i-j) = \Prob[P(t) = p_s+i-j]$, and the boundary terms reduce to the tail $\bar{F}_P(d) = \Prob[P(t) > d]$ and its complement. This recovers the standard form
\begin{equation}
    P_{ij} =
    \begin{cases}
        \bar{F}_P(p_s + i - 1), & j = 0,\\[2pt]
        p(p_s + i - j), & 1 \leq j \leq b - 1,\\[2pt]
        1 - \bar{F}_P(p_s + i - b), & j = b.
    \end{cases}
    \label{eq:transition-probs}
\end{equation}
For the Gaussian demand $P(t) \sim \mathcal{N}(\mu_P, \sigma_P^2)$, the probabilities in~\eqref{eq:mc-transition-general} are evaluated from the normal CDF $\Phi$. With $c_{ij} = p_s + (i-j)h$,
\begin{equation}
    P_{ij} =
    \begin{cases}
        1 - \Phi_P\big(p_s + b_i - h/2\big), & j = 0,\\[3pt]
        \Phi_P\big(c_{ij} + h/2\big) - \Phi_P\big(c_{ij} - h/2\big), & 0 < j < k, \\[3pt]
        \Phi_P\big(p_s + b_i - b + h/2\big), & j = k,
    \end{cases}
    \label{eq:mc-transition-gaussian}
\end{equation}
where $\Phi_P(x) \triangleq \Phi\!\big((x - \mu_P)/\sigma_P\big)$. The Gaussian density is integrated over the width-$h$ demand cell associated with each transition, and the two boundary states collect the tails that clamp the battery empty or full. The same construction applies to any demand distribution with a known CDF.

\subsection{Stationary overdraw probability}
\label{app:markov-overdraw}

The stationary distribution $\boldsymbol{\pi}$ solves $\boldsymbol{\pi}^{\!\top}\mathbf{P} = \boldsymbol{\pi}^{\!\top}$ with $\sum_i \pi_i = 1$. The overdraw probability is the stationary average of the per-state overdraw event $P(t) > p_s + b_i$,
\begin{equation}
    \Prob[\text{overdraw}] = \sum_{i=0}^{k} \pi_i \, \Prob \big[\,P(t) > p_s + b_i\,\big] = \sum_{i=0}^{k} \pi_i \, \bar{F}_P(p_s + b_i).
    \label{eq:mc-overdraw}
\end{equation}
The overdraw term uses the exact tail at each grid level, so the discretization affects only the dynamics of $b(t)$ and not the evaluation of the overdraw event. The required capacity $p_r(b)$ of~\eqref{eq:required-capacity} is then found by solving $\Prob[\text{overdraw}] = \varepsilon$ for $p_s$, for example by bisection.

%% file: prop_eff_pow.tex
\section{Effective Power and its Properties}
\label{sec:prop_eff_bw}

The effective power of a load characterized by the stochastic process $P(t)$ can be written as:
\begin{equation}
    \label{def:eff_bw-appendix}
    P_\text{eff}(r,T)=\frac{1}{rT} \log \mathbb{E}\left[ \exp \left(r\sum_{t=1}^T P(t) \right)\right],
\end{equation}
where $T$ is the time-scale of interest and $r$ is the transform variable for the moment-generating function of the process.

\begin{enumerate}
    \item Let the power level $P(t)$ take on values in $[0,p_{\max}]$. Effective power can be upper and lower bounded as:
    \[ \mu_P \leq P_\text{eff}(r,T) \leq p_{\max} \]
    for all pairs $(r,T)$.
    \item When statistically independent loads are multiplexed, their effective powers are additive: for $P_\text{sum}(t)=\sum_{k=1}^N P_k(t)$,
    \[P_\text{eff}^{(N)}(r,T) = \sum_{k=1}^N P_{\text{eff},k}(r,T). \]
    \item If $P(t)$ has uncorrelated samples (e.g., i.i.d.), then $P_\text{eff}(r,T)$ is constant in $T$:
    \[P_\text{eff}(r,T) = \frac{1}{r}\Lambda_P(r),\]
    where $\Lambda_P(r)$ is the log-moment generating function of $P(t)$. Note that if $P(t)<\infty$ with probability $1$, then $\Lambda_P(r)$ exists in some interval $(r^-,r^+)$ that includes the origin $r=0$.
    \item If $P(t)$ is ergodic, then
    \begin{equation}
    \label{eq:eff_bw_asymp}
    P_\text{eff}(r,\infty) \triangleq  \lim_{T\to \infty} P_\text{eff}(r,T) = \frac{1}{r}\bar{\Lambda}_P(r), \end{equation}
    where $\bar{\Lambda}_P(r)$ is the semi-invariant asymptotic log moment generating function of $P(t)$, which is a convex function of $r$ for the values for which the function is finite. For i.i.d. $P(t)$, $\bar{\Lambda}_P(r)=\Lambda_P(r)$. We refer to this definition of effective power in the core of our paper. 
\end{enumerate}
\noindent \textbf{Note:} For simplicity of notation, we will use $P_\text{eff}(r) \triangleq P_\text{eff}(r,\infty)$ for the effective power of processes with well-defined asymptotic log moment generating functions. Note that if the process is simply i.i.d., $P_\text{eff}(r)=P_\text{eff}(r,T)$ for all $T$. 

\noindent \textbf{Second order approximation:} For a load $P(t)$,
\begin{equation}
\label{eq:general_second_order}
P_\text{eff}(r,T) = \frac{1}{T} \mathbb{E}\left[ \sum_{t=1}^T P(t) \right] + \frac{r}{2T}\text{var}\left( \sum_{t=1}^T P(t) \right) + o(r^2).
\end{equation}
For an ergodic load,
\begin{equation}
\label{eq:ergodic_second_order}
P_\text{eff}(r,\infty) = \mu_P+\frac{1}{2} \sigma_\text{eff}^2 r + o(r^2),
\end{equation}
where $\mu_P$ is the sample mean\footnote{Since $P(t)$ is stationary, $\mu_P$ is also the $T$-step average} and $\sigma_\text{eff}^2$ was defined in Eq.~(\ref{eq:on-off-cov-1}).
The second order approximation becomes increasingly accurate as $r$ gets smaller. Note that the second order expansion\footnote{This expansion also covers situations with \textit{long-range dependent} power levels: $\text{var}\left( \sum_{t=1}^T P(t) \right) = \sigma^2 T^{2H}$, where $H\in(0,1)$ is the Hurst parameter. The Hurst parameter $H=\frac{1}{2}$ for i.i.d. and short-range dependent processes, including finite-state Markov-modulated stochastic processes.} is exact for Gaussian processes for all values of $r$.

%% file: typical_overdraw.tex
\section{Derivation of $T_{typ}(b)$}
\label{sec:T_typ}

Here, we will evaluate $T_\text{typ}(b)$ in the LBR and show that it grows linearly with $b$. Technically speaking, $\frac{T_\text{typ}}{b} \stackrel{\tiny{b\to \infty}}{\rightarrow}$ constant with probability $1$.

Let us define $\tau \triangleq \frac{T}{b}$ and rewrite the overflow probability:
\begin{align}
    \mathbb{P}[overdraw] &= \mathbb{P}\left(\sum_{t=-\tau b+1}^0 P(t) > p_s \tau b+b \right) \nonumber \\
    &= \mathbb{P}\left(\frac{1}{\tau b}\sum_{t=-\tau b+1}^0 (P(t) - p_s) > \frac{1}{\tau} \right) \nonumber \\
    &\leq \exp \left[ \tau b \left( \bar\Lambda_P(r) - \left( p_s-\frac{1}{\tau} \right)r \right) \right], \label{eq:chernoff_typ_time}
\end{align}
where (\ref{eq:chernoff_typ_time}) follows from the Chernoff bound. In the LBR, the tightest Chernoff exponent becomes tight and the typical duration becomes probabilistically dominant~\cite{dembo1998,tse1995multiplexing}, so we can write:
\begin{multline}
    \lim_{b\to \infty} \frac{1}{b} \log \mathbb{P}[overdraw] = \\ 
    \sup_{\tau>0}\left\{ \tau \inf_{r>0}\left( \bar\Lambda_P(r) - \left( p_s-\frac{1}{\tau} \right)r \right) \right\} .
    \label{eq:tightest_chernoff}
\end{multline}
The convexity of the asymptotic LMGF makes the optimization problem stated in (\ref{eq:tightest_chernoff}) amenable to a clean solution. Rather than applying the Karush-Kuhn-Tucker (KKT) conditions to the analytical expressions, the solution can be reached using geometric arguments that are fairly insightful.  

\begin{figure}[t]
    \centering
    \includegraphics[width=3in]{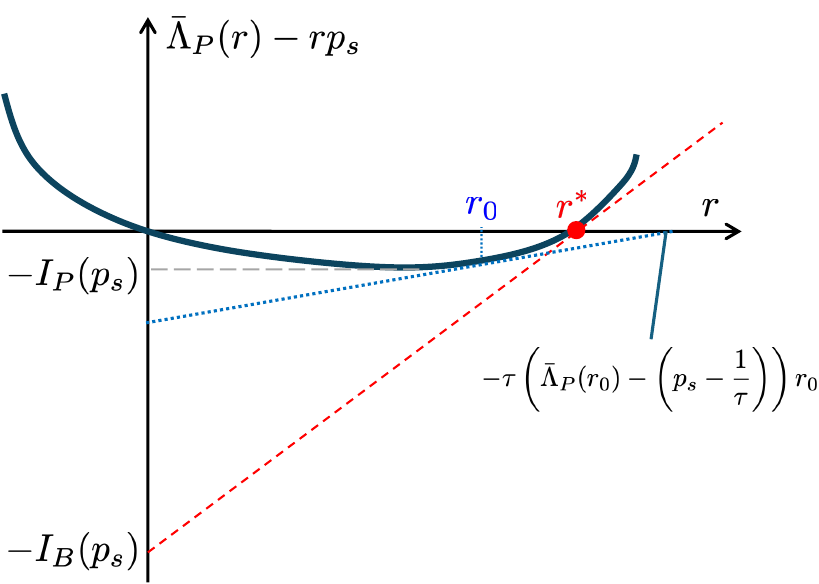}
    \caption{The semi-invariant log moment generating function of the per-user demand less per user power input. The minimization over $r$ is finding $r_0$ for a given $\tau$, which happens to be the abcissa of the tangent line to the curve. The maximization over $\tau$ is equivalent to minimizing the value of the point where the tangent line cuts the x-axis. The line cuts the y-axis at the point that gives the negative of the exact value of the rate function $I_B(p_s)$.}
    \label{fig:log_mmt}
\end{figure}

In Fig.~\ref{fig:log_mmt}, we illustrate $\bar\Lambda_P(r)-rp_s$ - the asymptotic LMGF of the difference between the power demand by a single unit (rack/server) and the per-user power supply ($p_s$). For a given $\tau$, the value of $r$ for the tightest Chernoff bound can be found by differentiating the exponent given in the right hand side (\ref{eq:tightest_chernoff}) with respect to $r$ and setting the result to $0$. Geometrically, the associated point is illustrated as $r_0$ in Fig.~\ref{fig:log_mmt}. When we extend the line that is tangent to the curve at $r_0$, the abcissa of the point at which the line cuts the x-axis turns out to be identical to exactly the negative of the exponent in (\ref{eq:tightest_chernoff}). So, the typical time is found by minimizing the value of that abcissa over all possible values of $\tau$. That point turns out to be the $r^*$ since all other values of $\tau$ would lead to slopes for the line leading to an exponent $r>r^*$. Also, the point the line cuts the y-axis is the rate function, $I_B(r)$, evaluated at $r=p_s$, the supply power.

After joint optimization, the probability of overflow can be written as:
\begin{equation}
\label{eq:rate_battery}    
\mathbb{P}[overdraw] \sim \exp(r^*b) = \exp (-\tau^* b I_B(p_s) ) .
\end{equation}
One can see here that $T_\text{typ}(b)=\tau^*b$, and since the slope of the curve at $r=r^*$ is identical to $\bar\Lambda_P'(r)-p_s,\ \tau^*$ can be found as the reciprocal of this slope. Hence,
\begin{equation}
    T_\text{typ}(b) = \frac{b}{\bar\Lambda_P'(r^*)-p_s}.
    \label{eq:typical_time_general-appendix}
\end{equation}
and $\mathbb{P}[overdraw] \sim \exp (-T_\text{typ}(b) I_B(p_s) )$. 

One final note; recall that for the NBR $\mathbb{P}[overdraw] = \exp (-I_P(p_s))$, where $I_P(p_s) = \sup_{r > 0} \big\{r p_s - \Lambda_P(r)\big\}$ was given in Eq.~\ref{eq:rate-function}. In Fig.~\ref{fig:log_mmt}, we also illustrate $I_P(p_s)$ and it can be seen that $I_P(p_s) < I_B(p_s)$.

%% file: app-region2a.tex
\section{Model of Region 2a}
\label{app:region2a}
This section approximates the slope of Region 2a for a symmetric ON/OFF process with state dwell time $L$, as in Section \ref{sec:on-off}. 

Consider a random OFF period of length $T_{\text{OFF}}$. In Region 2a, the battery is sufficiently small so that at the end of the OFF period, the battery has charged to $\min \{ p_sT_{\text{OFF}},b \} \approx b$, assuming it started the OFF period empty.  

Now consider the subsequent ON period of length $T_{\text{ON}}$ which starts with a full battery. Since in 2a the battery has already reduced the power to near the ON mean, $\mu_1$, we model the residual ON state variations as $0$ and thus the consumed power during this stage is $\approx \mu_1 T_{\text{ON}}$. Therefore, during this ON period, any additional battery storage can be used to reduce the required power by $\frac{b}{T_{\text{ON}}}$. Specifically, to avoid a overdraws, 
\begin{align}
p_s T_{\text{ON}} + b \geq u_1 T_{\text{ON}} \label{eqn:R2a-slope-condition}
\end{align}
Since $T_{\text{ON}}$ is geometrically distributed, $\mathbb{P}[T_{\text{ON}} > d] = \left(1 - \frac{1}{L}\right)^d$, and thus to keep (\ref{eqn:R2a-slope-condition}) from being violated with probability no more than $e^{-\delta}$, we require
\begin{align}
p_s \geq u_1 + s b 
\end{align}
where
\begin{align}
s =  \frac{1}{\delta_H} \log{\left(1 - \frac{1}{L} \right) } < 0
\end{align}
is the slope of the capacity reduction in Region 2a. For large $L$, $s \approx -\frac{1}{L \delta_H}$. Thus, longer dwell times and small target overdraws reduce the absolute value of the slope. 

Linear mean reduction can continue until the battery $b = b_{t,3}$, is too large to fully charge at the end of the OFF period,  when the system enters region 2b. Similar probabilistic arguments to the above can be made, however we empirically find that a simple average argument is sufficient to approximate the end of R2a.  Specifically, if the system spends $L$ amount of time in the OFF state, it can charge to at most $p_sL$ and this will be equal the battery size when $b_{t,3} = p_s L$. Now if the system is operating at margin, $p_s \approx u_1 + sb_{t,3}$, and assuming the size of R2a is approximated by $b_{t,3} - b_{t,2} \approx b_{t,3}$, then
\begin{equation}
b_{t,3} \approx \frac{\mu_1 L}{1 - sL}
\end{equation}

%% file: Appendix_setting_model_parameters.tex
\section{Setting Model Parameters from Empirical Power Traces}
\label{app:model_params}

To test our model against the empirical data, the following 4 step process was used to set the HIGH/LOW model parameters:   
\begin{enumerate}
    \item First, numerical determine the required power $p_r(b)$ for any desired overflow target $\varepsilon$ using the battery state equations (\ref{eq:battery-update}), as described above, and as shown in Fig.~\ref{fig:realdata-capacity}. 
    \item Second, set the HIGH state distribution by identifying the end of Phase 1. Specifically, identify the transition, $b_{\text{tx},2}$, between the first LBR region (R1b) and the mean reduction phase R2a. This transition is characterized by the transition from the diminishing slope of $p_r(b)$ in R1b and the constant slope in R2a. Given that, set $\mu_H = p_r(b_{\text{tx},2})$ based on the simplifying assumption that the battery has reduced the required power to the high state mean by the end of Phase 1.   Set the HIGH state variances using the Gaussian assumption and the empirical estimate of $p_r(0)$, i.e. from Eqn. (\ref{eqn:cap-sbr-mmp}), $\sigma_H = (p_r(0) - \mu_H )/\alpha_H$, where $\alpha_H = \sqrt{2\delta_H - \ln(4\pi\delta_H)}$  and $\delta_H = - \ln{(2\varepsilon)}$. Recall that the HIGH state can tolerate twice the average overdraw because the system spends half its time and the LOW state does not contribute to overdraws for battery sizes when the HIGH state dominates. 
    \item  Third, set the LOW state mean $u_L$ so that the overall process mean matches the empirical mean, i.e. $u_L = 2 \mu_P - \mu_H$, where $\mu_P$ is the real data mean. By assumption, $\sigma_L = \sigma_H$. 
    \item Third, set the dwell times $L = L_H = L_L$ from Eq.~(\ref{eqn:on-off-2a}) by calculating the slope $s$ of $p_r(b)$ in R2a. To do this, identify the transition, $b_{\text{tx},3}$, between the constant slope of R2a and the diminishing slope of R2b. Then set $s = (p_r(b_{\text{tx},3}) - p_r(b_{\text{tx},2})/(b_{\text{tx},3} - b_{\text{tx},2})$. 
\end{enumerate}

Note that the demand parameters are determined solely by 5 measurements: the process mean $\mu_P$, the no battery requirement, $p_r(0)$, and the start, end, and slope of R2a ($b_{\text{tx},2}$, $b_{\text{tx},3}$, and its slope $s$).  Referring to Fig. \ref{fig:realdata-capacity}, the empirical simulation of $p_r(b)$ is highly linear in R2a, making identification of its endpoints straightforward. Precise identification of the endpoints will not affect the overall parameters that much and our goal here is not to develop a generic algorithm to find R2a in any data set.

Note that identification of the transition between the SBR (R1a) and the start of the LBR (R1b), $b_{\text{tx},1}$, is not necessary to determine the model parameters.  For plots, we empirically set this transition to be consistent with the empirically settings of the other transition points. Thus, we set $b_{\text{tx},1}$ to the point where the slope of $p_r(b)$ becomes sublinear, as illustrated in the zoomed insets in Fig. \ref{fig:realdata-capacity}. An alternate approach would have been to set this transition point at the $50\%$ midpoint, $b_{\text{tx},1}  \approx 0.5 \alpha_H \sigma_H$, given the HIGH state gaussian i.i.d. assumption.